%% file: carina_rota.tex
\documentclass[preprint2]{aastex6}
\usepackage{natbib,twoopt}

\input{defs.tex}

\shortauthors{M. Fabrizio et al.}
\shorttitle{Carina Project. X. Kinematical properties.}

\begin{document}

\title{The Carina Project. X. On the kinematics\\of old and intermediate-age stellar 
populations\altaffilmark{1,2}.}
 
\author{M.~Fabrizio\altaffilmark{3,4},
G.~Bono\altaffilmark{5,6},
M.~Nonino\altaffilmark{7},
E.L.~{\L}okas\altaffilmark{8},
I.~Ferraro\altaffilmark{6},
G.~Iannicola\altaffilmark{6},\\
R.~Buonanno\altaffilmark{3,4},
S.~Cassisi\altaffilmark{4},
G.~Coppola\altaffilmark{9},
M. Dall'Ora\altaffilmark{9},
R.~Gilmozzi\altaffilmark{10},
M.~Marconi\altaffilmark{9},
M.~Monelli\altaffilmark{11,12},
M.~Romaniello\altaffilmark{10,13},
P.~B.~Stetson\altaffilmark{14},
F.~Th\'evenin\altaffilmark{15},
A.~R.~Walker\altaffilmark{16}
}
\email{fabrizio@oa-teramo.inaf.it}
\altaffiltext{1}{During the preparation of this manuscript, Giuseppina Coppola passed 
away. Her ideas, dedication and personality will be greatly missed.}
\altaffiltext{2}{Based on spectra either collected with VIMOS at ESO/VLT (090.D-0756(B),
P.I.: M.~Fabrizio), or retrieved from the ESO/ST-ECF Science Archive
Facility: 
FORS2 at ESO/VLT (072.D-0651(A), P.I.: G.~Bono; 078.B-0567(A),
P.I.: F.~Th\'evenin); \giraffe-UVES at ESO/VLT (074.B-0415(A),
076.B-0146(A), P.I.: E.~Tolstoy; 171.B-0520(A)(B)(C), 180.B-0806(B),
P.I.: G.~Gilmore); UVES at ESO/VLT (66.B-0320(A), P.I.: E.~Tolstoy).}
\altaffiltext{3}{ASI Science Data Center, Via del Politecnico s.n.c., I-00133 Rome, Italy}
\altaffiltext{4}{INAF-Osservatorio Astronomico di Teramo, Via Mentore Maggini s.n.c., I-64100 Teramo, Italy}
\altaffiltext{5}{Dipartimento di Fisica, Universit\`a di Roma "Tor Vergata", Via della Ricerca Scientifica 1, I-00133 Roma, Italy}
\altaffiltext{6}{INAF-Osservatorio Astronomico di Roma, Via Frascati 33, I-00040 Monte Porzio Catone (RM), Italy}
\altaffiltext{7}{INAF-Osservatorio Astronomico di Trieste, Via G.B. Tiepolo 11, I-40131 Trieste, Italy}
\altaffiltext{8}{Nicolaus Copernicus Astronomical Center, Bartycka 18, 00-716 Warsaw, Poland}
\altaffiltext{9}{INAF-Osservatorio Astronomico di Capodimonte, Salita Moiariello 16, I-80131 Napoli, Italy}
\altaffiltext{10}{European Southern Observatory, Karl-Schwarzschild-Str. 2, G-85748 Garching bei M\"unchen, Germany}
\altaffiltext{11}{Instituto de Astrof\'{i}sica de Canarias, Calle Via Lactea s/n, E-38200 La Laguna, Tenerife, Spain}
\altaffiltext{12}{Departamento de Astrof\'{i}sica, Universidad de La Laguna, E-38200 La Laguna, Tenerife, Spain}
\altaffiltext{13}{Excellence Cluster Universe, Boltzmannstr. 2, D-85748, Garching bei M\"unchen, Germany.}
\altaffiltext{14}{Dominion Astrophysical Observatory, NRC-Herzberg, National Research Council, 5071 West Saanich Road, Victoria, BC V9E 2E7, Canada}
\altaffiltext{15}{Universit\'{e} de Nice Sophia-antipolis, CNRS, Observatoire de la C\^{o}te d'Azur, Laboratoire Lagrange, BP 4229, F-06304 Nice, France}
\altaffiltext{16}{Cerro Tololo Inter-American Observatory, National Optical Astronomy Observatory, Casilla 603, La Serena, Chile}

\begin{abstract}
We present new radial velocity (RV) measurements of old (horizontal
branch) and intermediate-age (red clump) stellar tracers in the Carina
dwarf spheroidal. They are based on more than 2,200 low-resolution
spectra collected with VIMOS at VLT. The targets are faint
(20$\lesssim$\vv$\lesssim $21.5~mag), but the accuracy at the faintest
limit is $\le$9~\kms. These data were complemented with RV measurements
either based on spectra collected with FORS2 and \giraffe\ at VLT or
available in the literature. We ended up with a sample of 2748
stars and among them 1389 are candidate Carina stars. We found that the
intermediate-age stellar component shows a well defined rotational
pattern around the minor axis. The western and the eastern side of the
galaxy differ by +5 and --4~\kms\ when compared with the main RV peak.
The old stellar component is characterized by a larger RV dispersion and
does not show evidence of RV pattern. 
We compared the observed RV distribution with $N$-body simulations for a
former disky dwarf galaxy orbiting a giant Milky Way-like galaxy
\citep{lokas15}. We rotated the simulated galaxy by 60 degrees with
respect to the major axis, we kept the observer on orbital plane of the
dwarf and extracted a sample of stars similar to the observed one.
Observed and predicted $V_{rot}/\sigma$ ratios across the central
regions are in remarkable agreement. This evidence indicates that Carina
was a disky dwarf galaxy that experienced several strong tidal
interactions with the Milky Way. Owing to these interactions, Carina
transformed from a disky to a prolate spheroid and the rotational
velocity transformed into random motions. 
\end{abstract}

\keywords{galaxies: dwarf --- galaxies: individual (Carina) --- galaxies: stellar content --- galaxies: kinematics and dynamics}

\maketitle

\section{Introduction}
\label{sec:intro}
Dwarf galaxies are fundamental laboratories for cosmological, chemical evolution, 
synthetic stellar population and evolutionary models. These systems have been the 
cross-road of paramount photometric and spectroscopic investigations aimed at 
constraining their star formation history, chemical evolution and 
dark matter (DM) content. 
Multi-band photometry based either on ground-based mosaic CCD cameras
\citep{carney86,mateo91,smecker96,majewski00,monelli03,tolstoy09,bono10,stetson11} 
or on space images \citep{monelli10tucana,monelli10cetus,hidalgo13,skillman14}
provided the opportunity to identify the faint Turn-Off in several dwarf
galaxies in the Local Volume and to perform robust separation of galaxy
stars from both background galaxies and foreground stars using the
color-color plane in several Local Group galaxies.
The kinematic investigations typically lag when compared with the photometry. 

However, during the last few years the use of multi-slit and multi-object spectrographs 
provided the opportunity to measure the radial velocity (RV) of stellar samples ranging 
from a few hundreds (WLM, \citealt{leaman13wlm}) to a few thousands (Fornax, 
\citealt{walker11}). These measurements are crucial to estimate the dynamical mass, and 
in turn to constrain the mass-to-light ratio of these intriguing systems. However, it is not 
clear yet whether dwarf irregulars are --at fixed total mass-- less DM dominated than 
dwarf spheroidal (dSph, \citealt{woo08,sanna10}). 

Recent findings also suggest that the well studied dSphs appear to have, within a factor of 
two, an universal mass profile \citep{walker09b,strigari08,mateo93}. Moreover, the 
increase in the sample size and in the RV precision provided the opportunity to identify 
kinematical substructures in several nearby dwarf galaxies. They have already been 
detected in Sextans \citep{walker08}, LeoI \citep{mateo08}, Fornax and Sculptor 
\citep{walker11} and in Carina \citep{fabrizio11}.
   
The reason why we are interested in dSphs is threefold.

{\em (a) ---} Empirical evidence indicates that dSphs and ultra-faint dwarfs 
(UFDs) are the smallest stellar systems to be DM dominated. Therefore, 
they can provide firm constraints on the smallest DM halos that can retain 
baryons. 

{\em (b) ---} Cosmological models suggest that dSphs are the fossil records 
of the Galactic halo. Therefore, their kinematic and chemical properties 
can provide firm constraints on the formation and evolution of the Milky Way.  

{\em (c) ---} Current cosmological models indicate that the density profile 
of DM halos is a steep power law towards the center (cusp-profile, \citealt{navarro97}). 
On the other hand, tentative empirical evidence indicates that several 
dSphs and Low-Surface-Brightness galaxies disclose a constant density 
profile (core profile, \citealt{kormendy04}). 
The key observables to settle this open problem 
is the rotation curve. The cusp model has a rotation curve that increases 
as the square root of the radius, whereas the core model has a rotation 
curve that increases linearly with radius (see also \citealt{deblok10}). 
Carina is a perfect laboratory 
to address the formation and evolution of gas-poor dwarf galaxies. 

This is the tenth paper of a series focussed on the stellar populations of 
the Carina dSph. We have investigated the evolutionary and pulsation 
properties of stellar populations in Carina using homogeneous and accurate 
optical photometry, metallicity distributions and kinematics. The structure 
of the paper is as follows. In Sect.~\ref{sec:data} we present the different 
spectroscopic data sets adopted in the current investigation. Special attention 
was paid to data reduction and calibration of the new low-resolution 
VIMOS spectra and to the observing strategy (\S~\ref{subsec:calib}). 
The approach adopted 
to measure radial velocities and the analysis of the radial velocity 
distributions are discussed in Sect.~\ref{sec:rv}. Sect.~\ref{sec:photo} deals 
with the photometric index \cubi\ adopted to separate old and intermediate-age 
stellar populations along the Carina red giant branch (RGB). 
The radial velocity maps of the Carina sub-populations are considered in 
Sect.~\ref{sec:rvmap}, while the variation (radial and angular) of the velocity 
profiles are addressed in Sect.~\ref{sec:rvvar}. In Sect.~\ref{sec:simul} we perform 
a detailed comparison between current radial velocity distributions and recent 
$N$-body simulations provided by \citet{lokas15}. 
Finally, Sect~\ref{sec:summary} gives a summary of the current results 
together with a few remarks concerning the future of the Carina project.


\begin{figure*}
\centering
\includegraphics[width=0.75\textwidth]{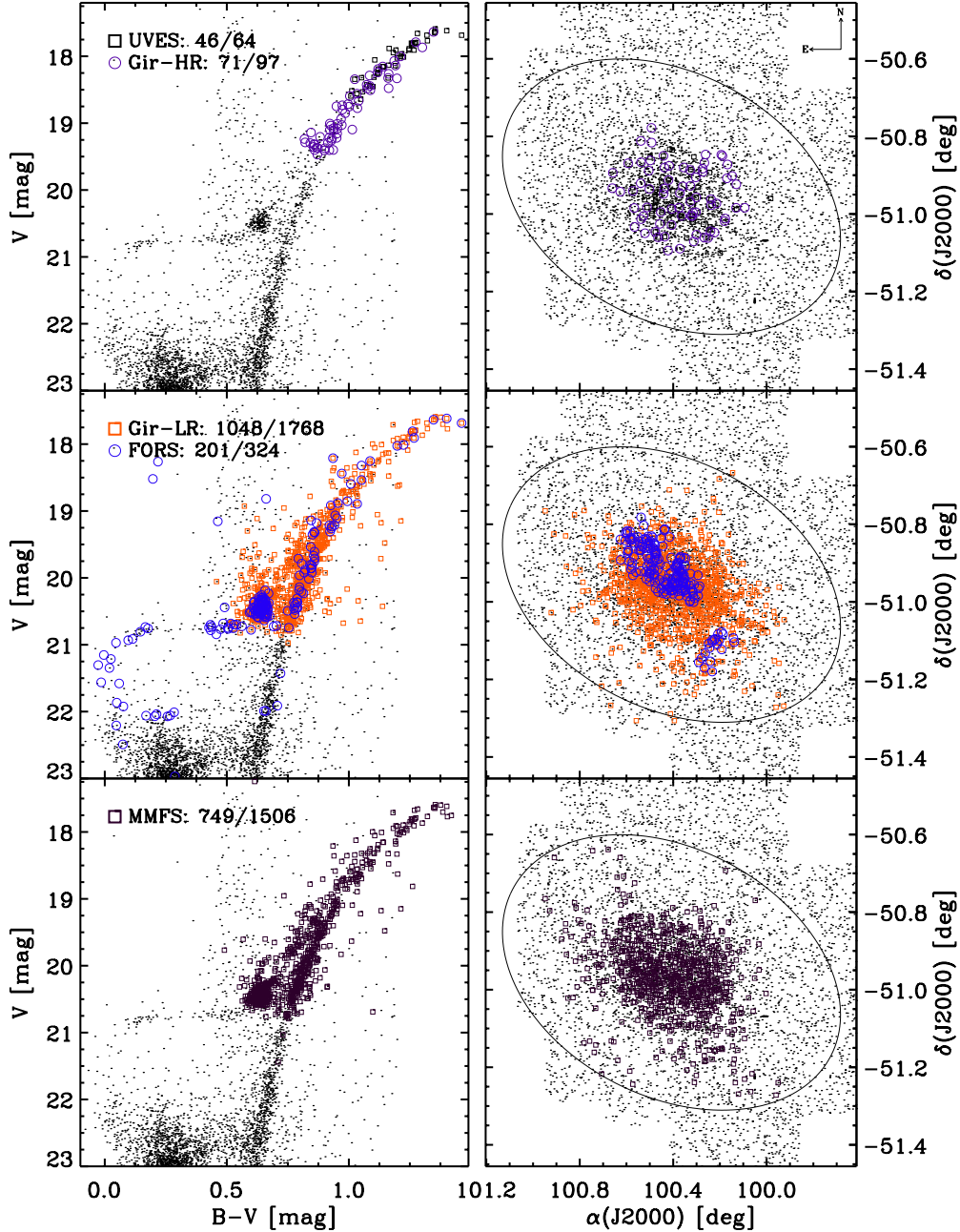}
\caption{Left panels---distribution of the spectroscopic targets in the 
\vv\ vs \bmv\ color-magnitude diagram. 
From top to bottom targets with RV measurements based on 
high-resolution and medium-resolution spectra or available in 
the literature. 
Right panels---radial distribution of the spectroscopic targets overplotted 
on our Carina photometric catalog (black dots). The ellipse shows the Carina 
truncation radius.\label{cmd}}  
\end{figure*}

\begin{figure*}
\centering
\includegraphics[width=0.75\textwidth]{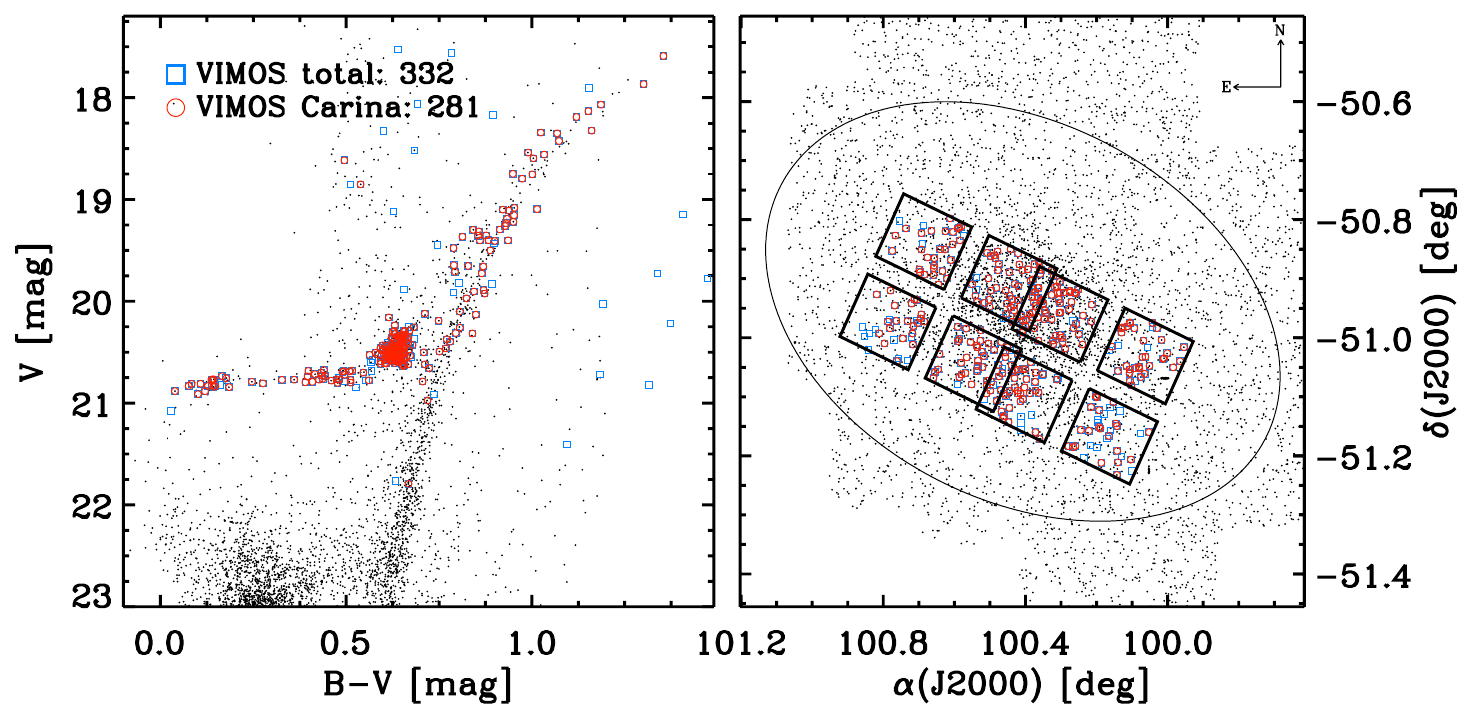}
\caption{Same as Fig.~\ref{cmd}, but for the spectroscopic targets observed with VIMOS. 
Blue squares and red circles display candidate field and Carina stars. The black 
squares plotted in the right panel display the two VIMOS pointings (four chips each). 
\label{cmdVIMOS}}  
\end{figure*}

\section{Spectroscopic data sets and data reduction}
\label{sec:data}

To constrain the Carina kinematical structure, we adopted data collected 
at four spectrographs mounted at the Very Large Telescope (VLT) of the 
European Southern Observatory (ESO). The spectral resolution of data ranges 
from 600 to 40,000, allowing us to obtain accurate RVs up to the 
horizontal branch (HB) magnitude 
level (\vv$\sim$21.5 mag). The overlap between different data sets allowed us 
to provide solid constraints on possible systematic errors. The adopted 
data sets and the sample of Carina stars observed are the following: 

{\em High Resolution}---  
The high spectral resolution sample includes 79 spectra for 64 stars collected 
with the Ultraviolet and Visual Echelle Spectrographs 
\citep[UVES, R$\sim$40,000;][]{dekker00} 
either in fiber (FLAMES/UVES) or in slit (UVES) mode. This data set also includes 
1983 spectra for 97 stars collected with three high-resolution grisms 
(HR10, HR13, HR14A) of the Fibre Large Array Multi Element Spectrograph 
multi-fiber spectrograph \giraffe\ 
\citep[GHR, R$\sim$20,000;][]{pasquini02}. 
As a whole this data set mainly includes RG stars brighter than \vv$\sim$19.5~mag.  
The typical accuracy of the radial velocity measurements based on these spectra is 
better than 0.8~\kms\ for the entire sample \citep{fabrizio11,fabrizio12}. 
The color-magnitude diagram (CMD) and 
the sky distribution of HR targets are shown in top panels of Fig.~\ref{cmd}.
 
{\em Medium Resolution}--- 
This data set includes more than $1.6\times10^4$ spectra for 1768 stars 
collected with the \giraffe\ multi-fiber spectrograph (GMR, R$\sim$6,500) using the 
grating LR08 centered on the NIR calcium triplet (middle panels of Fig.~\ref{cmd}). 
These spectra were complemented with 772 low-resolution (R$\sim$2,500,
grisms: 1028z, 1400V) spectra for 324 stars collected with the multi-slit 
FOcal Reducer/low dispersion Spectrograph 2 (FORS2, \citealt{appenzeller98}).
The typical accuracy of the radial velocity measurements based on these spectra is 
better than 6-10~\kms\ for the entire sample \citep{fabrizio11}.  
The key advantage of the current spectroscopic sample is that it covers the 
entire body of the galaxy also covering the regions across the tidal radius 
(r$_t$=28.8$\pm$3.6\arcmin, \citealt{mateo98araa}) and beyond ($\sim$0.7\deg). 
Moreover and even more importantly, this sample covers a broad range in magnitudes 
(17.5$\lesssim$\vv$\lesssim$23 mag). This means a good overlap with HR spectra 
(60 objects), and the opportunity to trace different stellar populations: 
old, low-mass, helium burning stars ($t\sim 12$~Gyr, Horizontal Branch stars); 
intermediate-age, helium burning stars ($t\sim 6$~Gyr, Red Clump [RC] stars);       
red giants (RG) that are a mix of old and intermediate-age stars
(see the middle panels of Fig.~\ref{cmd}).

Note that both HR and MR spectra have already been discussed by \citet{fabrizio11} 
and \citet{fabrizio12} to constrain the kinematical properties of the Carina dSph. 
The reader interested in a more detailed discussion concerning the data reduction 
and the calibration of the different instruments is referred to the quoted 
papers. 

{\em Literature}---
We collected radial velocity measurements for 1506 stars provided by \citet{walker07} 
using the high-resolution (20,000-25,000) spectra collected with 
Michigan\slash MIKE fiber system (MMFS) at Magellan (see the bottom panels of 
Fig.~\ref{cmd}). 
The 909 objects in common allowed us to estimate possible systematics and we found 
that they are quite similar, with a difference of $\approx$0.7~\kms\ and a standard 
deviation of 11~\kms.      

{\em Low Resolution}---  
This data set includes 2454 spectra for 332 stars collected with the multi-slit
VIsible MultiObject Spectrograph \citep[VIMOS, R$\sim$600;][]{lefevre03} using the 
MR-GG475 grism (4,800$\le$$\lambda$$\le$10,000~\AA) during Dec. 2012 and 
Jan. 2013\footnote{090.D-0756(B), P.I.: M.~Fabrizio}. 
This is a new data set and it is presented here for the first time.
The spectra were collected with VIMOS in service mode and Table~\ref{tab:logobs} 
gives the log of the observations. The spectra were collected in good seeing 
conditions, and indeed, it is on average better than 1\arcsec\ (see column 6) 
and with an airmass smaller than 1.2\arcsec. The spectra were collected with two 
pointings (see columns 2, 3 and 4) covering a substantial fraction of the body of the 
galaxy. 
The field of view of VIMOS is made by four quadrants of 7\arcmin$\times$8\arcmin\ 
each, separated by a cross shaped gap that is $\sim$2\arcmin\ wide. Special 
attention was pain to designing the masks of the individual pointings. We split 
them into deep and shallow masks. The former ones host in the four quadrants more 
than $\sim$150 slits, 0.8\arcsec\ wide. The bulk of the targets of the deep masks 
are HB and RC stars. To improve the SNR of the faint targets the same masks were 
repeated seven times, moreover, 
the two pointings partially overlap across the center 
of the galaxy (see the right panel of Fig.~\ref{cmdVIMOS}). To improve the spatial 
sampling, the shallow masks cover the same sky area, the bulk of the targets 
are RG stars, and they typically host $\sim$110 slits, with similar slit width. 
These targets are brighter and the masks have only been repeated three times.   

The faint sample covers a limited range in magnitudes 
(20$\lesssim$\vv$\lesssim$21.5 mag). This means the 
opportunity to properly trace old (blue and red HB stars) and intermediate-age 
(RC stars) stellar populations. These spectra also have a good overlap with both 
HR (15 objects) and MR (178 objects) spectra. 
The left panel of Fig.~\ref{cmdVIMOS} 
shows the distribution in the CMD of both the candidate field stars 
(cyan squares) and the candidate Carina stars (red circles) according to 
the kinematic selection. The right panel of the same figure shows the
eight VIMOS pointings and the radial distribution of candidate field and 
galaxy stars. 

\input{tab_logobs.tex}

\begin{figure*}
\centering
\includegraphics[width=0.75\textwidth]{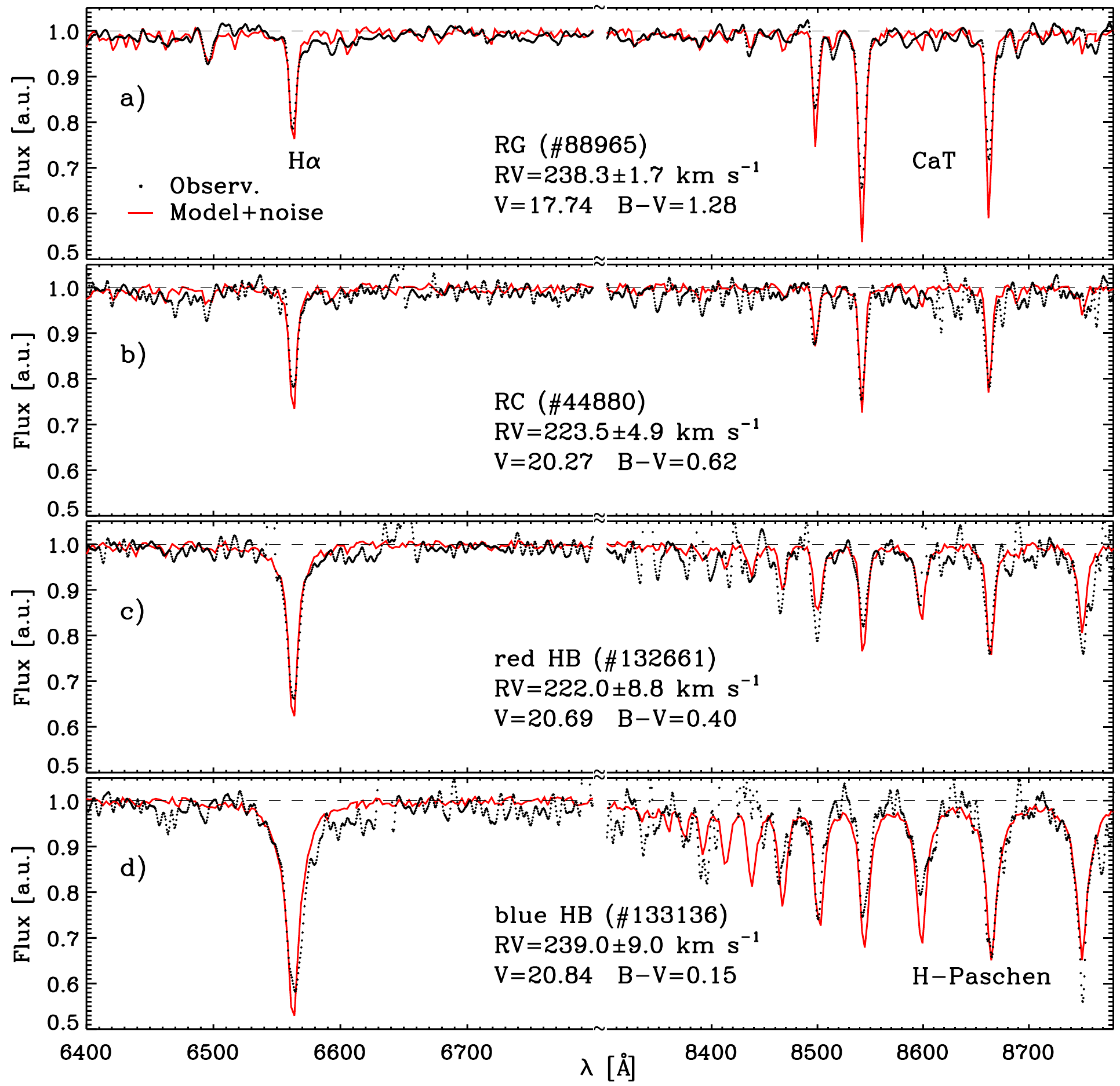}
\caption{Panel a)---Normalized VIMOS spectrum for a Red Giant (RG) star 
centered around the H$\alpha$ (left) and the \caii\ triplet (right). 
The black dots display the observed spectrum, while the red line 
the synthetic spectrum. The number in parentheses show the ID of the 
star in the photometric catalog. The radial velocity and its uncertainty 
is also labeled together with the visual magnitude and the \bmv\ color 
of the target.  
Panel b)---Same as Panel a), but for a Red Clump (RC) star. 
Panel c)---Same as Panel a), but for a red HB star. Note that in the 
right portion of the spectrum the region around the hydrogen 
Paschen lines is shown.  
Panel d)---Same as Panel c), but for a blue HB star. See text for more 
details. 
\label{spectra}}  
\end{figure*}

\begin{figure*}
\centering
\includegraphics[width=0.75\textwidth]{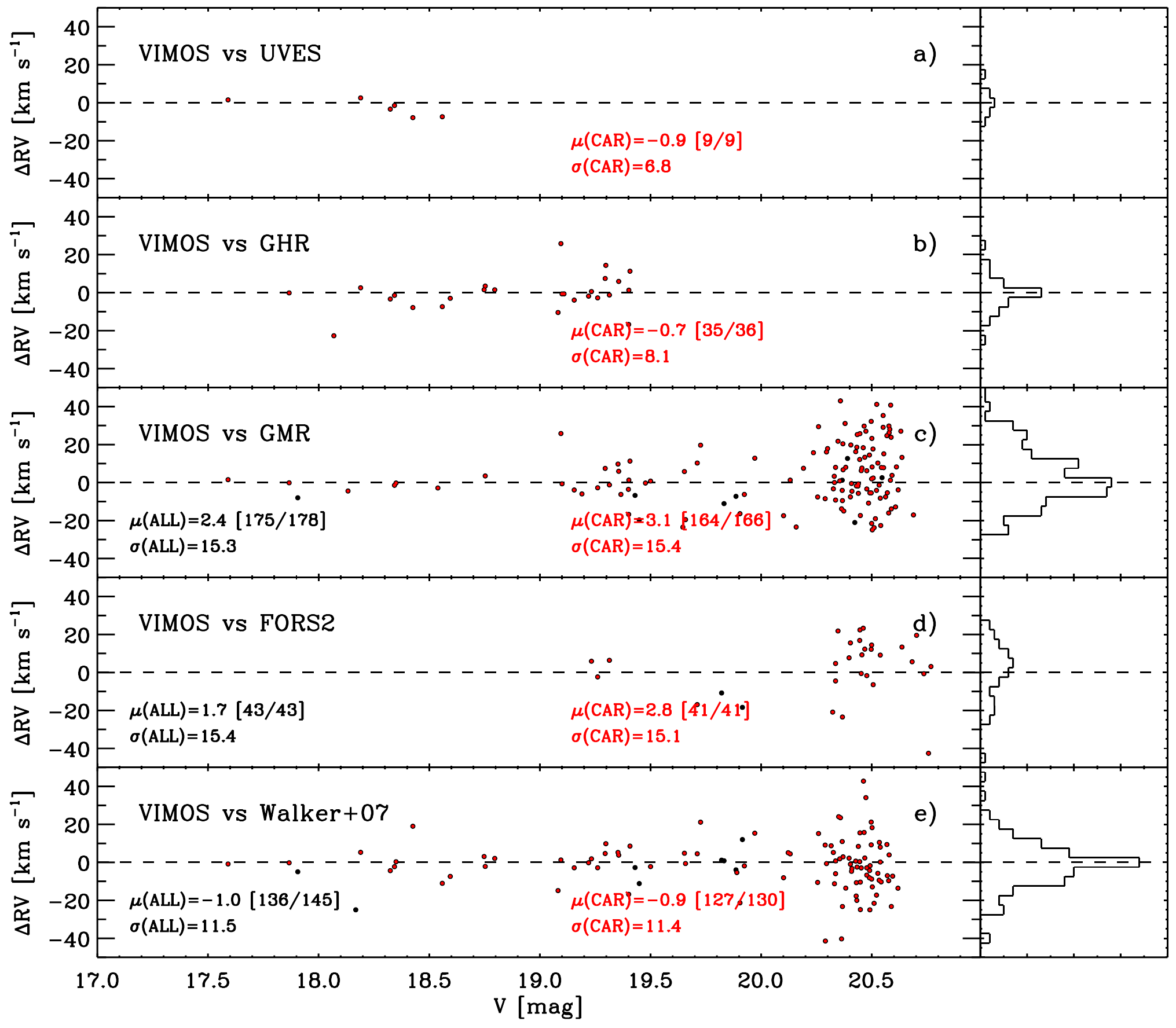}
\caption{Comparison between RV measurements based on low-resolution spectra 
collected with VIMOS and literature estimates. 
The biweight mean ($\mu$), the standard deviation for the entire sample, and for the 
candidate Carina stars (180$\le$RV$\le$260~\kms) are labeled.
The numbers in brackets show the number of objects in common between the two 
samples before and after the biweight mean. \label{errRV}}  
\end{figure*}

\subsection{VIMOS spectra: extraction and wavelength calibration}
\label{subsec:calib}

The observed spectra were individually extracted using the following approach. 
We wrote a custom software in R\footnote{\tt https://cran.r-project.org} to 
extract the slits data from the bias corrected spectra, with the associated 
flat and wavelamp. The slit sky model was constructed via a row by row robust 
polynomial fit and then subtracted from the slit. The sky subtracted slits 
for the same target were combined using {\tt imcombine} in 
IRAF\footnote{IRAF is distributed by the National Optical Astronomy Observatory, 
which is operated by the Association of Universities for Research in Astronomy, 
Inc., under cooperative agreement with the National Science Foundation.}. The 
ensuing reference spectrum was extracted from the combined images using 
{\tt apall} in IRAF. 
This reference spectrum was used to extract the single slit spectrum, which 
in turn, was also used as a reference in the {\tt apall} procedure to extract 
the associated sky and the associated wavelength lamp. The wavelength 
calibration of the spectra was performed using {\tt identify-reidentify} 
IRAF tasks. Subsequently, every sky spectrum was cross-correlated with 
a synthetic sky radiance spectrum computed using 
SkyCalc\footnote{\tt www.eso.org/observing/etc/skycalc/skycalc.htm}, 
and the shift in wavelength was applied to the target spectra. 
To overcome possible systematics on radial velocity measurements caused by 
non optimal centering of the star in the slit, we cross-correlated the star 
spectra with a synthetic sky transmittance spectrum over the wavelength 
range 7550-7750~\AA, and the shift in wavelength applied accordingly.
In particular, we used telluric absorption lines in the Fraunhofer A band
(7167-7320~\AA, 7593-7690~\AA) to correct for possible star miscentering
in slit (e.g., \citealt{sohn07}). A synthetic transmittance spectrum is
cross correlated with the star spectrum, in the wavelength range of the 
Fraunhofer A band, essentially the star is used as a lighthouse with the 
incoming light affected by the Earth also through atmospheric telluric 
absorption. The telluric feature positions and line shapes are 
pressure and wind dependent (e.g., \citealt{griffin73}), however, they cause 
variations of the orders of tens of \ms, i.e., well below the errors in
our observations.

To validate the precision of the approach adopted to reduce and calibrate   
low-resolution spectra, the black dots plotted in Figure~\ref{spectra} show 
the VIMOS spectra of four stars ranging from a relatively bright Red Giant 
star (RG, \vv=17.74~mag, panel a) to a fainter Red Clump star 
(RC, \vv=20.27~mag, panel b) and to even fainter red and blue HB 
stars (RHB, BHB, panels c and d) located at \vv=20.69 and \vv=20.84~mag.
Note that the quoted figure shows the wavelength regions located around the 
strong H$\alpha$ line ($\lambda$=6562.8~\AA) and the regions around the 
\caii\ triplet for cool and warm targets and the H-Paschen lines for the 
hotter RHB and BHB stars. This means that we can fully exploit the 
large wavelength range covered by VIMOS spectra in dealing with stars 
covering a broad range of effective temperatures and surface gravities.  

To further constrain the plausibility of the current approach 
Figure~\ref{spectra} shows also the comparison with synthetic spectra 
(red solid lines, see Sect.~\ref{sec:rv}). To improve the precision of 
the RV velocity measurements, the synthetic spectra were degraded 
to the VIMOS spectral resolution using a Gaussian kernel and 
added noise to reach the typical signal-to-noise ratio of observed spectra.
The main outcome is that the accuracy of 
the RV measurements based on VIMOS spectra range from better than 2~\kms\
for the bright sources to better than 9~\kms\ for the faintest sources 
in our sample.  
This evidence becomes even more compelling if we account for the total 
exposure time adopted, namely 7.5 hours per pointing and each pointing 
including, on average, 260 targets. 

\section{Radial velocity measurements}
\label{sec:rv}

The radial velocity measurements were performed for each target on individual 
spectra, instead of measuring the radial velocity of the co-added spectrum. 
The latter was the approach adopted in \citet{fabrizio11,fabrizio12rossa} 
to determine the radial velocity of both high-, medium- and low-resolution 
spectra. The current measurements are based on a new procedure performing an 
automatic cross-match between the observed spectrum and a synthetic one. 
The change in the data analysis strategy was motivated by the fact that the 
low-resolution spectra collected with VIMOS (R$\sim$600) have a limited 
number of isolated and un-blended spectral features that can be safely adopted 
to provide accurate and precise radial velocity measurements. The adopted 
synthetic spectra were extracted from the Pollux 
Database\footnote{\tt pollux.graal.univ-montp2.fr} \citep{palacios10}, with 
stellar parameters of typical RGB, RC, red HB and blue HB stars. This means that 
we selected a grid of models with the following input parameters: 
\teff=[5000, 5500, 6500, 7500]~K, \logg=[0.5, 2.5, 2.5, 2.5]~dex, a stellar 
mass equal to 1~M$_\sun$ and spherical geometry. Moreover, the above models 
where computed at fixed chemical composition, \feh=--1.50, and \afe=+0.40. 
The reader interested in detailed discussion concerning the
iron and the $\alpha$-element distribution of old- and intermediate-age 
sub-populations is referred to \citet{fabrizio15}. The models were 
degraded to VIMOS spectral resolution using a Gaussian filter with the 
typical signal-to-noise ratio of observed data ($\sim$10-50). We performed 
a number of tests using models covering a broader range in input parameters
and we found that the radial velocity measurements are minimally affected by 
plausible changes ($\Delta$\teff=300 K, $\Delta$\logg=0.2~dex, 
$\Delta$\feh=0.2) in the adopted synthetic spectra.   

Specifically the automatic procedure performs a Least Squares minimization 
of the residual between observed spectrum and synthetic spectra. We performed 
a number of tests with medium- and high-resolution spectra degraded at the 
spectral resolution of VIMOS spectra and we found that the typical precision 
of the current radial velocity measurements is systematically better than 
12~\kms\ for both cool and warm targets.

\input{tab_RV.tex}

\subsection{Validation of radial velocity measurements}
\label{subsec:valid}

In order to validate the new radial velocity measurements based on VIMOS 
spectra, we performed a series of comparisons with radial velocities 
measured on medium- and high-resolution spectra. The left side of panel a) 
in Fig.~\ref{errRV} shows the comparison between VIMOS measurements and those 
based on high-resolution spectra (nine stars common) 
as function of the visual magnitude. The right side of the same panel displays 
the distribution of the residuals. The panel b) displays similar comparison and 
distribution as in panel a), but for the 36 GHR stars in common. Although the number 
of objects in common is smaller than four dozen and relatively bright, the residuals are 
quite small and symmetric. The mean (biweight, \citealt{beers90}) difference is 
$<$0.7~\kms, while the standard deviation is 7-8~\kms.

Panels c) and d) display the difference with radial velocities based on 
medium-resolution spectra (GMR, FORS2). The VIMOS sample has 178 stars 
in common with the GMR sample (black dots) and among them 166 stars are, 
according to the selection criteria discussed 
in Sect.~\ref{subsec:rvdistr}, candidate Carina 
stars (red dots). The objects in common with FORS2 are 43 and among them 
41 are candidate Carina stars. In this case, the mean difference is, on 
average of the order of 2~\kms,  while the standard deviation is 
$\sim$15~\kms\ (see labelled values for the different sub-samples).

The last panel e) shows the comparison between radial 
velocity measurements based on VIMOS spectra and similar measurements 
available in the literature from \citet{walker07}.
The quoted sample has a large number of objects in common, 145 stars, 
ranging from the tip of the red giant branch down to red clump stars. 
The mean difference is, once again, minimal ($<$1~\kms) and the residuals are 
symmetrical. Thus suggesting the lack of systematics in radial velocities 
based on low-resolution VIMOS spectra. 

It is noteworthy that the procedure we devised to calibrate VIMOS data, 
as described in Sect.~\ref{subsec:calib}, has substantially improved and 
reduced the differences between VIMOS radial velocities with similar 
measurements, but based on higher resolution spectra (see Fig.~\ref{errRV}). 
In particular, the quoted differences increase by a factor of ten, 
and the dispersion increases by a factor of three, if the wavelength 
calibration of VIMOS spectra was only based on lamps. The improvement 
mainly relies on the procedure adopted to extract and to correct for 
slit centering, i.e. by performing the cross-match with radiance and 
transmittance sky spectra, respectively.
\begin{figure*}
\centering
\includegraphics[width=0.75\textwidth]{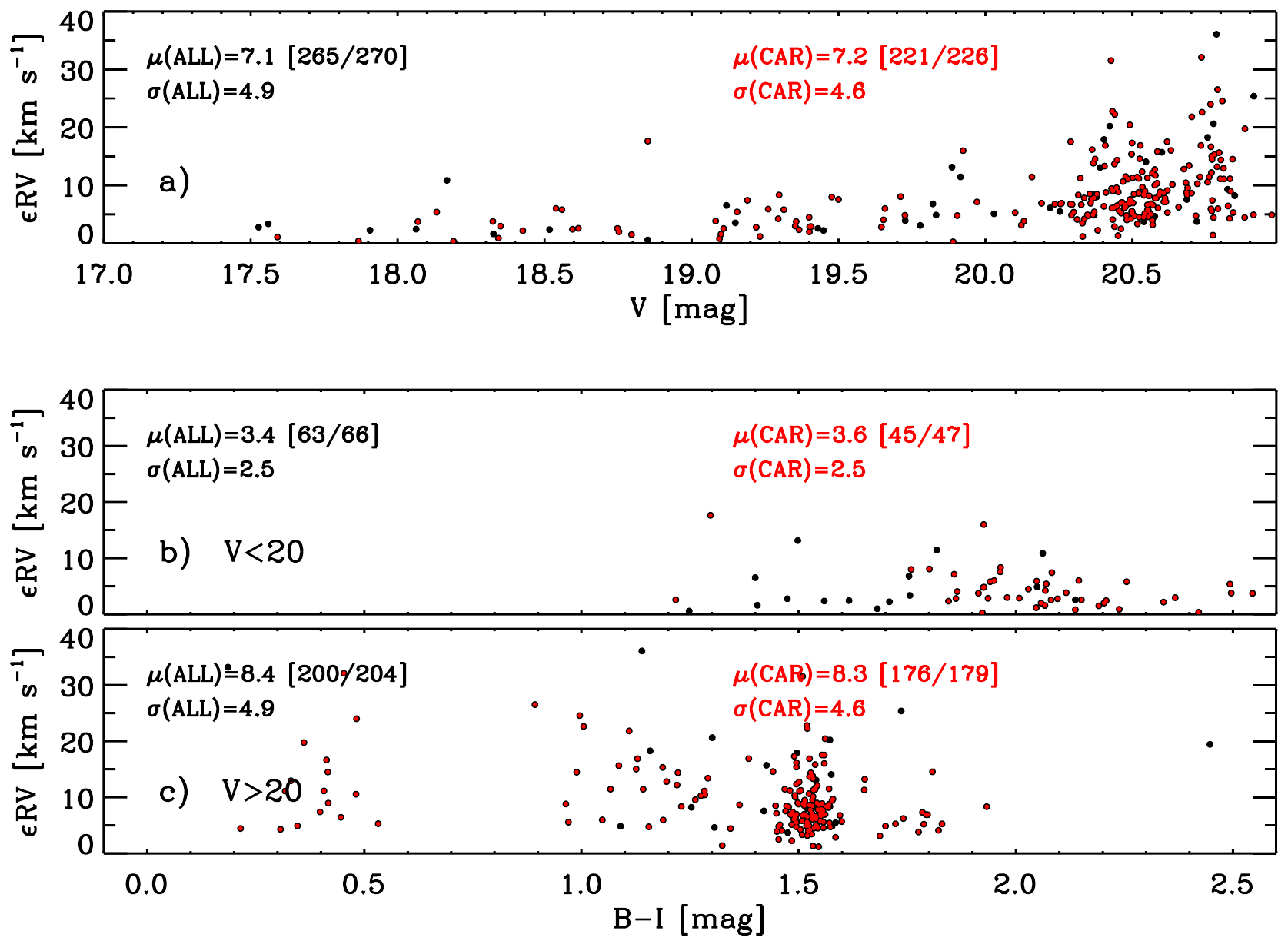}
\caption{Radial velocity uncertainties ($\epsilon$RV) as function of \vv\ magnitude 
(panel a)) and \bmi\ color (panels b) and c)). In particular, panel b) shows targets with
\vv$<$20~mag (mainly RGB stars), while panel c) displays targets with \vv$>$20~mag 
(HB and RC). The labels and the numbers in parentheses have the same meaning 
as in Fig.~\ref{errRV}.}
\label{eRV}
\end{figure*}

To further validate the RV measurements based on VIMOS spectra, and in 
particular the possible occurrence of systematics as a function of 
magnitude and spectral type, Fig.~\ref{eRV} shows the RV errors ($\epsilon$RV) 
as function of both \vv\ magnitude (panel a) and \bmi\ color (panels b) and c)). 
Note that to overcome the color overlap between different stellar sub-groups,  
the panel b) mainly shows RGB stars brighter than \vv$\sim$20~mag, while 
the panel c) mainly shows HB and RC stars fainter than \vv$\sim$20~mag. 

Data plotted in the quoted panels show no evidence of systematics in 
RV errors as function of magnitude (panel a)), with a mean RV error 
for candidate Carina stars of 7.2~\kms\ and a velocity dispersion of 4.6~\kms\ 
and the typical steady increase in the faint limit. The same outcome 
applies to the RV errors as a function of color (panels b) and c)), since
their distribution, within the errors, is similar for blue 
(\bmi$<$0.4~mag) and red (\bmi$>$1.5~mag) spectroscopic 
targets. Moreover, the above results apply, as expected, to both 
field and galaxy stars. These finding allow us to end up with a robust 
and homogeneous catalog of RV measurements.

Table~\ref{tab:RV} lists the radial velocity measurements of the
entire spectroscopic sample. Columns 1-3 give the identity and
coordinates of the stars, while columns 4-6 give the magnitudes in \vv,
\bb, \ii\ and relative errors (from \citealt{bono10, stetson11}). The 
mean RVs are listed in column 7 with their measurement errors, while 
column 8 gives, for each, target the spectrographs adopted to collect the 
spectra. In the last column is given a stellar population flag according 
to the \vv\ vs \cubi\ plane and to the CMD (see Fig.\ref{photosel}).

\input{tab_gauss.tex}

The occurrence of binary stars among RG stars in dSph galaxies is a
highly debated topic and their impact on the RV dispersion ranges 
from a sizable effect \citep{queloz95} to a modest error
\citep{olszewski96,hargreaves96} when compared with the statistical
error. However, in a recent investigation \citet{minor10} found, by
using a detailed statistical approach, that dSph galaxies with RV
dispersions ranging from 4 to 10 \kms\ can be inflated by no more than
20\%\ due to the orbital motion of binary stars. The key advantage of 
the current sample \citep{fabrizio11} is that a significant fraction of 
stars in our sample have spectra collected on a time interval of 
more than ten years. Therefore, the
current RV measurements are less prone to significant changes caused by
binary stars. Moreover, the conclusions of the current investigation are
minimally affected by a possible uncertainty of the order of 20\%\ in the
RV dispersion.

\begin{figure*}
\centering
\includegraphics[width=0.75\textwidth]{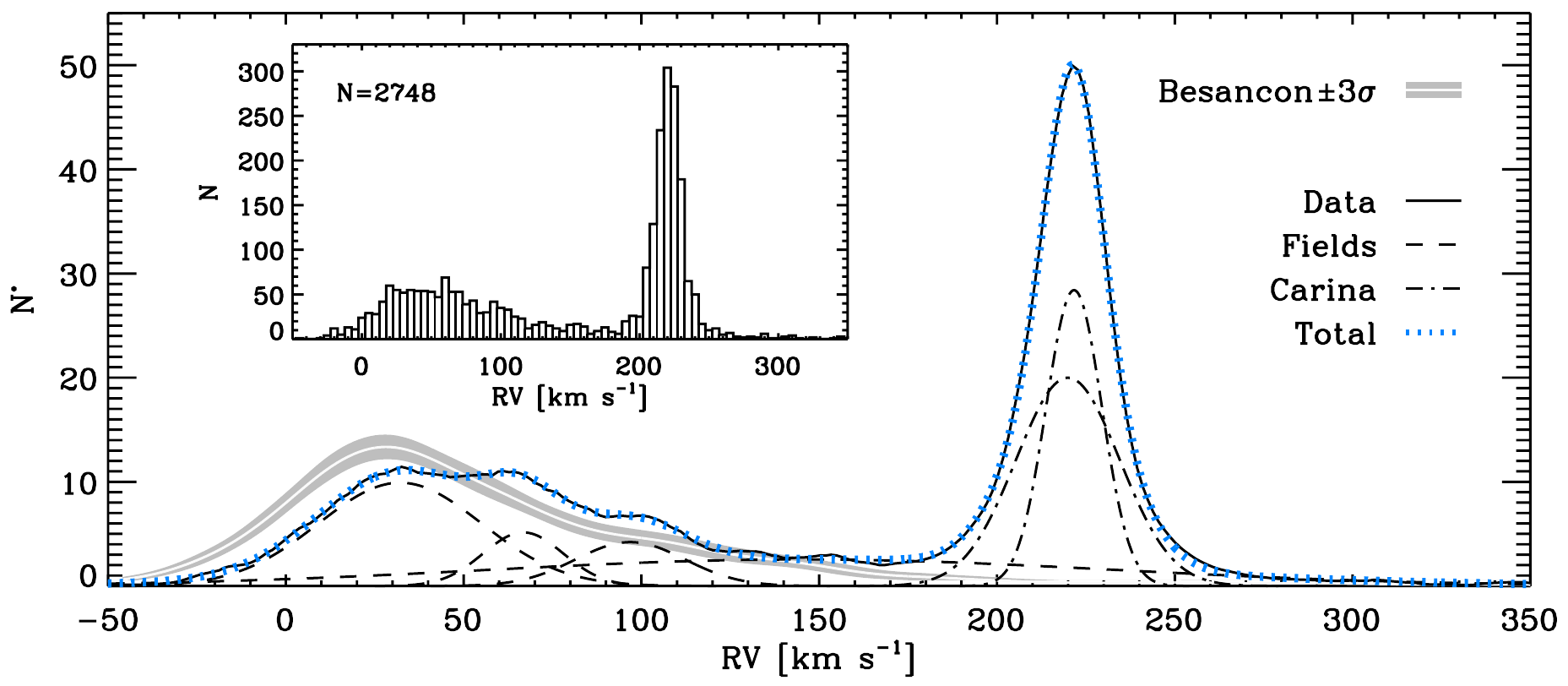}
\caption{Global radial velocity distribution of the entire spectroscopic sample.
The inset shows the histogram of the radial velocity measurements. 
The black solid line shows the smoothed radial velocity distribution using the 
Gaussian kernel. The dashed curves display the four Gaussians adopted to fit 
the candidate field stars, while the dashed-dotted curves show the two Gaussians 
adopted to fit the candidate Carina stars. The light blue dotted line shows the 
sum of the six Gaussians. 
The grey shaded curve shows the predicted mean radial velocity 
distribution, with the 3$\sigma$ error bar, based on the Besan\c{c}on Galaxy
model \citep[see the text for more details,][]{robin03}.
\label{RVdistr}}  
\end{figure*}

\begin{figure*}
\centering
\includegraphics[width=0.75\textwidth]{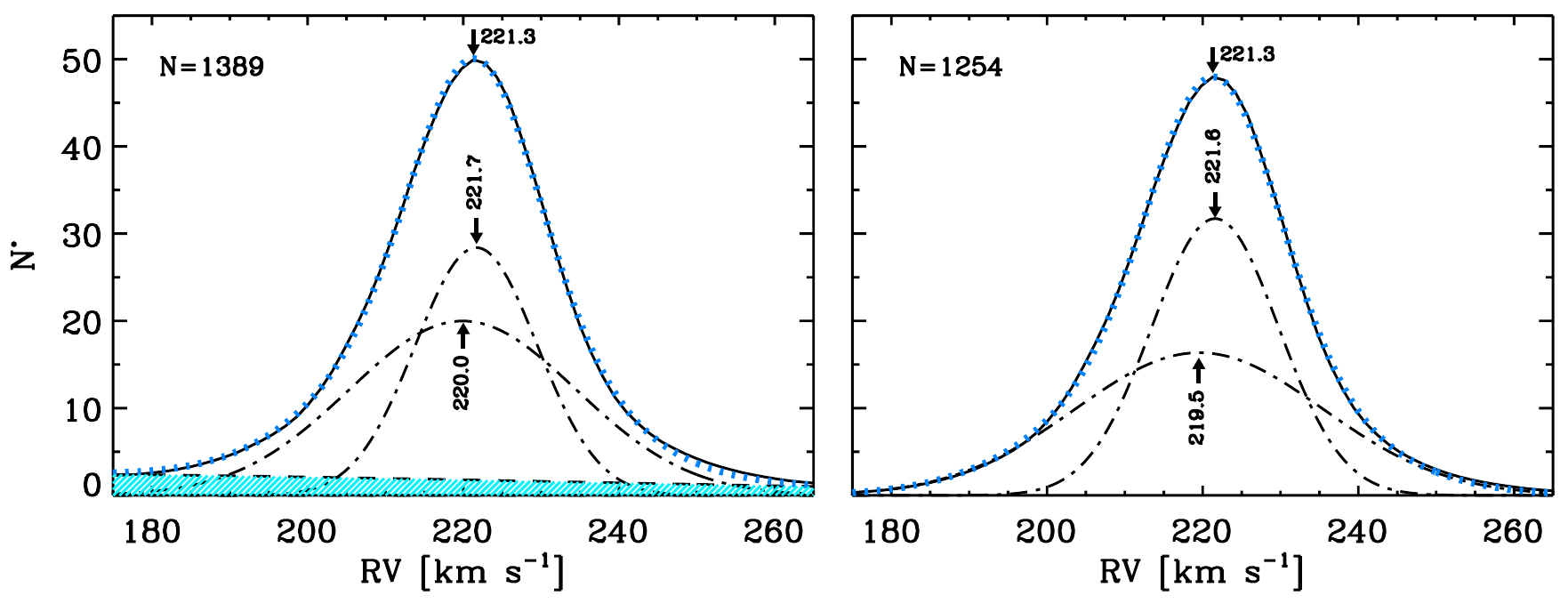}
\caption{
Left---Smoothed radial velocity distribution of 
candidate Carina stars (black curve). The dashed-dotted curves display 
the two Gaussians adopted to fit the RV distribution, while the light blue 
dotted curve their sum. The three arrows mark the peaks in radial velocity 
and the values are also labelled (see also Tab.~\ref{tab:multigauss}).
The cyan area discloses the possible overlap between candidate field and 
Carina stars.
Right---Same as the left panel, but after the random cleaning of field star 
contaminants (black curve). See text for more details. 
\label{RVdistrzoom}}  
\end{figure*}

\subsection{Radial velocity distribution}
\label{subsec:rvdistr}

On the basis of our spectra (HR, MR, LR spectra) we ended up with homogeneous radial 
velocity measurements for 2112 stars. The entire sample of RV measurements 
(our plus literature) includes 2748 stars. We applied a $4\sigma$ selection criterium 
(180$<$RV$<$260~\kms) and ended up with 1389 RV measurements of candidate Carina 
stars. 
The histogram plotted in the inset of Fig.~\ref{RVdistr} shows the radial 
velocity distribution 
of candidate field and Carina stars. To overcome subtle problems in constraining their 
RV distribution caused by the binning of the data, we associated to each star a Gaussian 
kernel with a $\sigma$ equal to its intrinsic error on the RV measurement. 
The black solid curve was computed by summing all the Gaussians over the entire 
RV range.
A glance at the quoted distribution shows a clear separation between candidate field 
and Carina stars. However, the criterium adopted for selecting candidate Carina stars, 
only based on the $\sigma$ of the RV distribution, might be affected by a possible bias. 
The high radial velocity tail of candidate field stars might attain RVs similar to the 
truly Carina stars. To overcome this problem we performed a global fit of the 
radial velocity distribution using six Gaussians. The number and the positions 
of the Gaussians are arbitrary and they were selected to minimize the residuals 
over the entire velocity range. Among them, four were used to fit the RV distribution 
of candidate field stars and two for the RV distribution of candidate Carina stars. 
Table~\ref{tab:multigauss} lists the adopted parameters for the multi-Gaussian fits. 

In order to test the plausibility of the separation between field and galaxy 
stars, the anonymous referee suggested to compare the current radial velocity 
distribution with the Besan\c{c}on Galaxy 
model\footnote{\tt http://model.obs-besancon.fr} \citep{robin03}. 
We randomly extracted from a complete simulation of 0.45 square degrees around the 
Carina coordinates a number of stars similar to our sample of candidate field stars 
($\sim$1360). 
The same extraction was repeated 100 times to take account of the limited area 
covered by the current spectroscopic sample and we computed their mean radial 
velocity distribution and their standard deviation. Predictions from the 
Besan\c{c}on Galaxy model were plotted as a gray shaded curve in Fig.~\ref{RVdistr}. 
The agreement between theory and observations is quite good over the entire 
velocity range of field stars (--20$\lesssim$RV$\lesssim$180~\kms). 
There is a mild excess 
of predicted stars in the low radial velocity tail (RV$\lesssim$30~\kms), but this 
appears as a minor problem, since we did not apply any local normalization.  
On the other hand, the predicted high tail vanishes for radial velocities 
larger than 150~\kms. Thus suggesting that we are minimally overestimating 
the possible contamination of field stars within the main Carina peak. However, 
the above differences are within the current theoretical and empirical uncertainties 
\citep{czekaj14}.
       
\begin{figure*}
\centering
\includegraphics[width=0.8\textwidth]{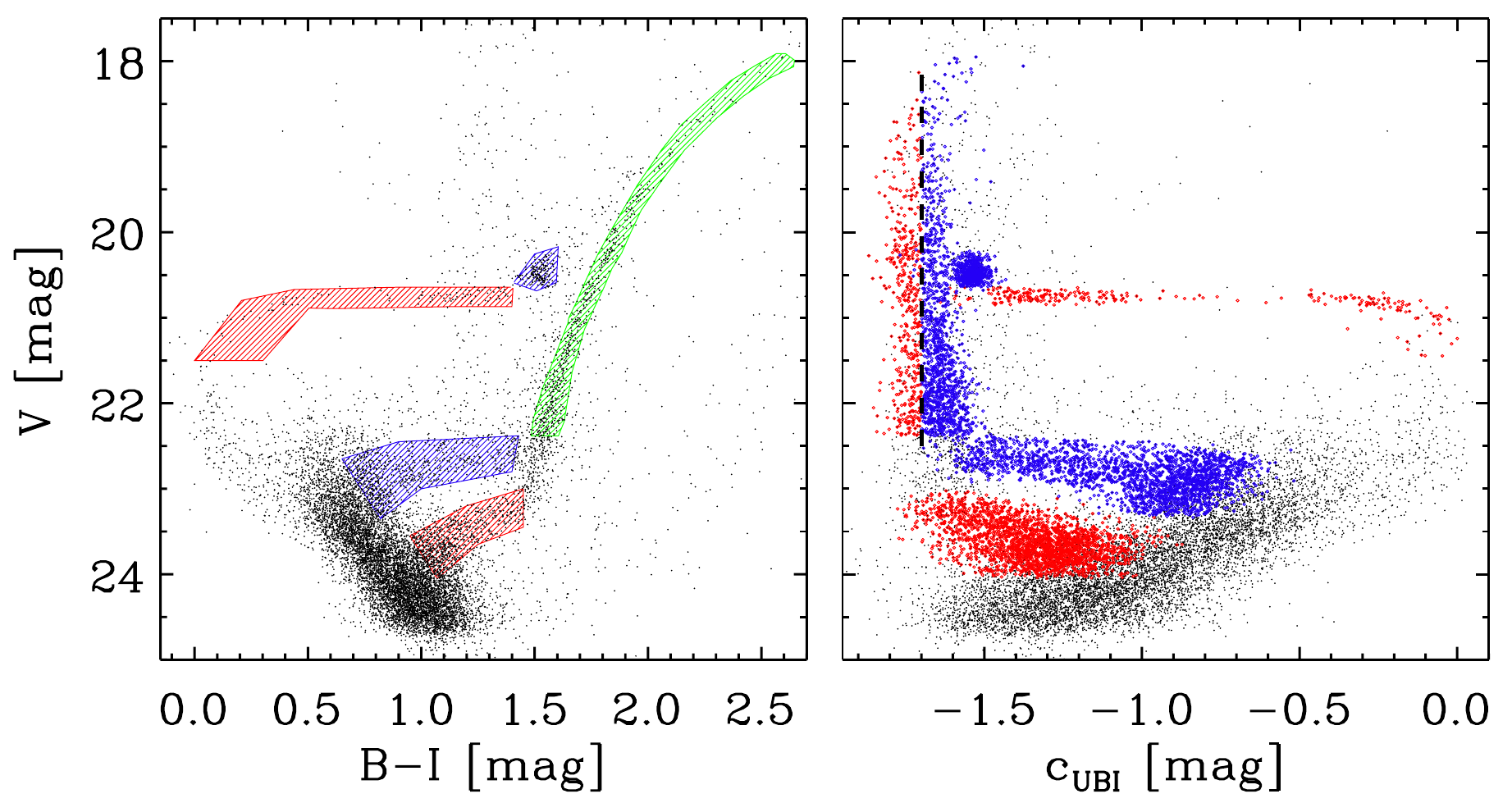}
\caption{
Left -- Photometric selection of old (red) and intermediate-age (blue) 
candidate Carina stars based on the \vv\ vs \bmi\ CMD. The former one included HB stars 
and faint sub-giant branch stars, while the latter RC stars and bright sub-giant branch 
stars. The green area shows RGB stars. 
Right -- Same as the left panel, but in the  \vv\ vs \cubi\ diagram. The vertical black 
dashed line shows the separation along the RGB between candidate old (red) and 
intermediate-age (blue) 
Carina stars. The old and the intermediate-age candidates selected in the left panel are 
also plotted with red and blue symbols.\label{photosel}}  
\end{figure*}

\begin{figure*}
\centering
\includegraphics[width=0.98\textwidth]{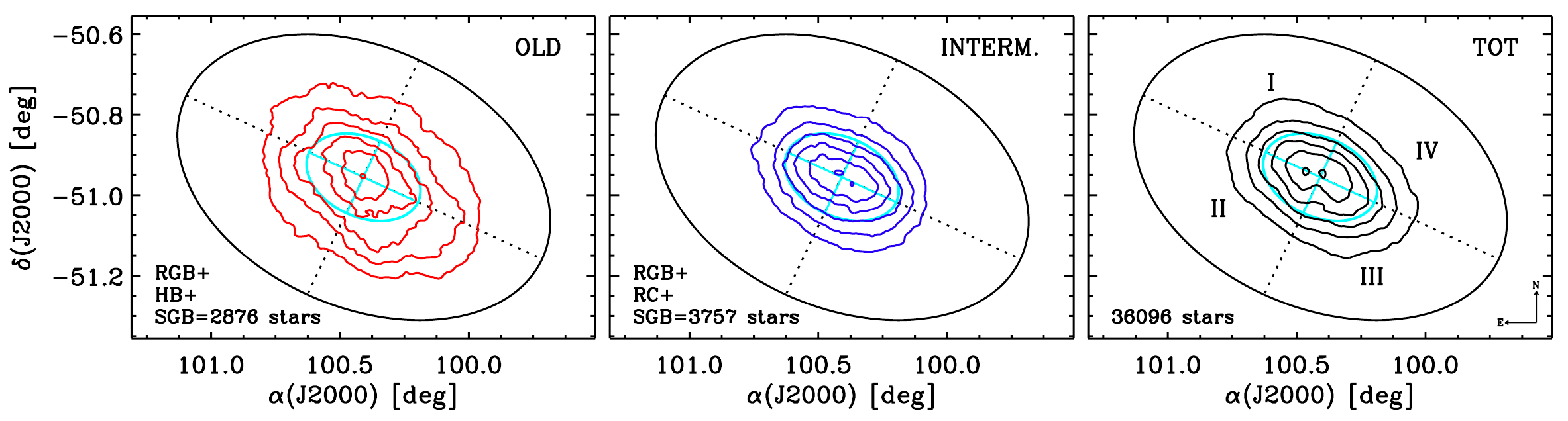}
\caption{Iso-contours of old (left, $\sim$2880 stars) and intermediate-age 
(middle, $\sim$3760 stars) stellar populations, selected using the photometric 
criteria showed in Fig.~\ref{photosel}. The black contours plotted in the right 
panel display the spatial distribution of the entire sample (36100) of candidate 
Carina stars (see, e.g., \citealt{bono10}). The cyan ellipse shows the Carina core radius.
The adopted numbering for quadrants is also labelled.
\label{contour}}  
\end{figure*}

To further constrain the quoted contamination, the left panel of Fig.~\ref{RVdistrzoom} 
shows a zoom of the RV distribution across the Carina peak. The asymmetry in the Carina 
RV distribution has already been discussed by \citet{fabrizio11}. Here we only mention 
that the main peak and the $\sigma$ attain values (220.74$\pm$0.28~\kms, 
$\sigma$=10.5~\kms) that agree quite well with previous estimates. Moreover, the 
individual Gaussians (see also Table~\ref{tab:multigauss}) clearly indicate that the broad 
component is mainly responsible of the asymmetry in the RV distribution, while the narrow 
one is mainly shaping the RV peak. Once again, the number and the analytical functions 
adopted to perform the fit of the main peak are arbitrary and they were selected to 
minimize the residuals. 
The cyan area plotted in the bottom of the RV distribution shows the possible 
contamination of candidate Carina stars. We estimated a contamination of 135 field stars 
over a sample of 1389 candidate Carina stars. They were randomly extracted ending up 
with a catalog of 1254 candidate Carina stars.   

\input{tab_mean.tex}
	
The right panel of Fig.~\ref{RVdistrzoom} shows the RV distribution of 
the candidate Carina stars after the subtraction of possible contaminants. 
We performed a series of Monte Carlo simulations in which we randomly extracted 
135 contaminants among their candidate Carina stars. The RV distribution plotted 
in Fig.~\ref{RVdistrzoom} is the mean over 100 independent extractions. Note 
that both the global fit and the individual Gaussians, due to the sample size 
and radial velocity precision, are minimally affected by the inclusion/exclusion 
of the possible contaminating field stars.      
The parameters of the final Gaussian fit are also listed in Table~\ref{tab:multigauss} 
together with their uncertainties. The uncertainties are the standard deviations of the 
fits over the 100 samples of cleaned RV distributions. 
In passing, we note that the broad and the narrow Gaussians display a difference 
in the peak of 1~\kms, while the $\sigma$ of the former one is almost a factor 
of two larger than the latter one. This suggests that the two quoted components 
appear to trace stellar populations characterized by different kinematic 
properties.  

In passing, we note that in a detailed 
photometric---based on DDO photometry---and spectroscopic
investigation, \citet{majewski05} found evidence of possible
contaminants from the Large Magellanic Cloud at radial distances of the
order of few degrees from the Carina center. We did not check for these
foreground stars, since the current kinematic investigation is limited
to stars located inside the Carina tidal radius. We plan to perform a 
detailed search for foreground stars and for extra-tidal Carina stars 
\citep{munoz06} in a forthcoming investigation (Fabrizio et al. 2016, 
in preparation).

\begin{figure}
\centering
\includegraphics[width=0.45\textwidth]{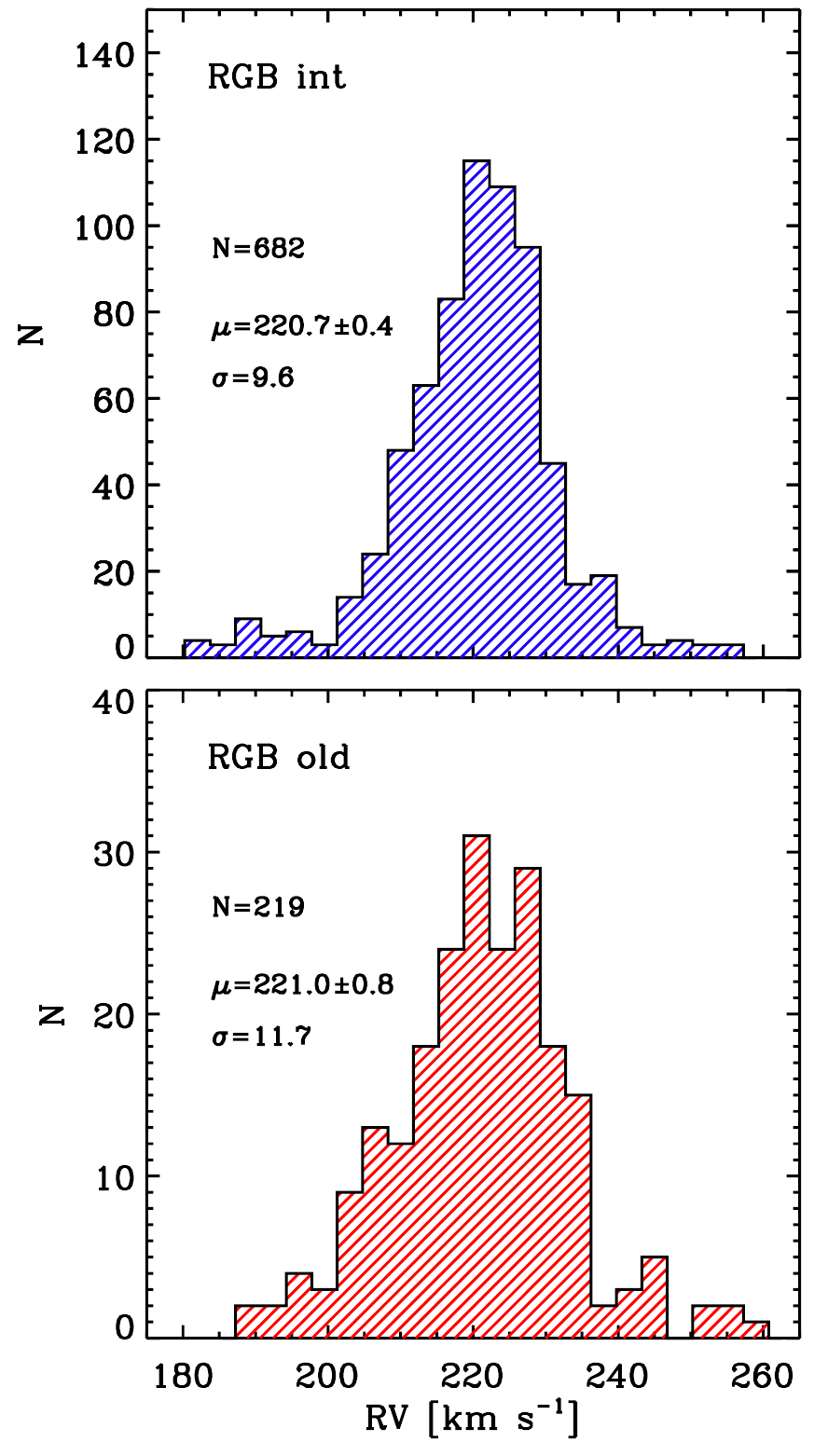}
\caption{Top---RV distribution of candidate 
intermediate-age RGB stars. The number of stars, biweight mean (with error) 
and standard deviation values are also labelled. 
Bottom---Same as the top panel, but the red shaded area displays candidate old 
RGB stars.
\label{hist_rgb}}  
\end{figure}

\section{Photometric selection}
\label{sec:photo}

The optical \vv, \bmi\ color-magnitude diagram of Carina offers the 
opportunity to easily identify several evolutionary phases that 
can be associated either with the old (HB, subgiant branch [SGB], plotted 
in red in the left panel of Fig.~\ref{photosel}) or the intermediate-age 
(RC, SGB, plotted in blue in the left panel of Fig.~\ref{photosel}).  
The two SGBs are too faint for the current 8-10 class telescopes. 
This is the reason why we are only left with 74 HB and 414 RC stars 
as solid bright tracers of the old- and of the intermediate-age 
sub-populations.
To further increase the sample of old- and intermediate-age 
stars we took advantage of the new photometric 
index--- \cubi=(\umb)--(\bmi) ---introduced by \citet{monelli13}. 
This index allow us, for the first time, to separate the two 
main sub-populations of Carina (old, intermediate-age) 
along a significant portion of the RGB \citep{monelli14}. 

The right panel of Figure~\ref{photosel} shows the Carina brightest portion 
in the \vv, \cubi\ diagram. The colored symbols display the kinematical 
selected stars and the different colors are related to different evolutionary 
phases and ages. The black dashed line shows the edge between old (red) and 
intermediate-age (blue) RGB stars, and its location was fixed following 
\citet{monelli14}. 
The red and blue symbols mark also the 
positions of HB (\vv=20.75, \cubi=$-1.5/0$ mag)  and RC (\vv=20.5, 
\cubi=$-1.6/-1.5$~mag) stars, respectively. The iso-contour of sky distributions for 
these two stellar components, and of the total sample as well, 
are shown in Fig.~\ref{contour}.
\begin{figure*}
\centering
\includegraphics[width=0.98\textwidth]{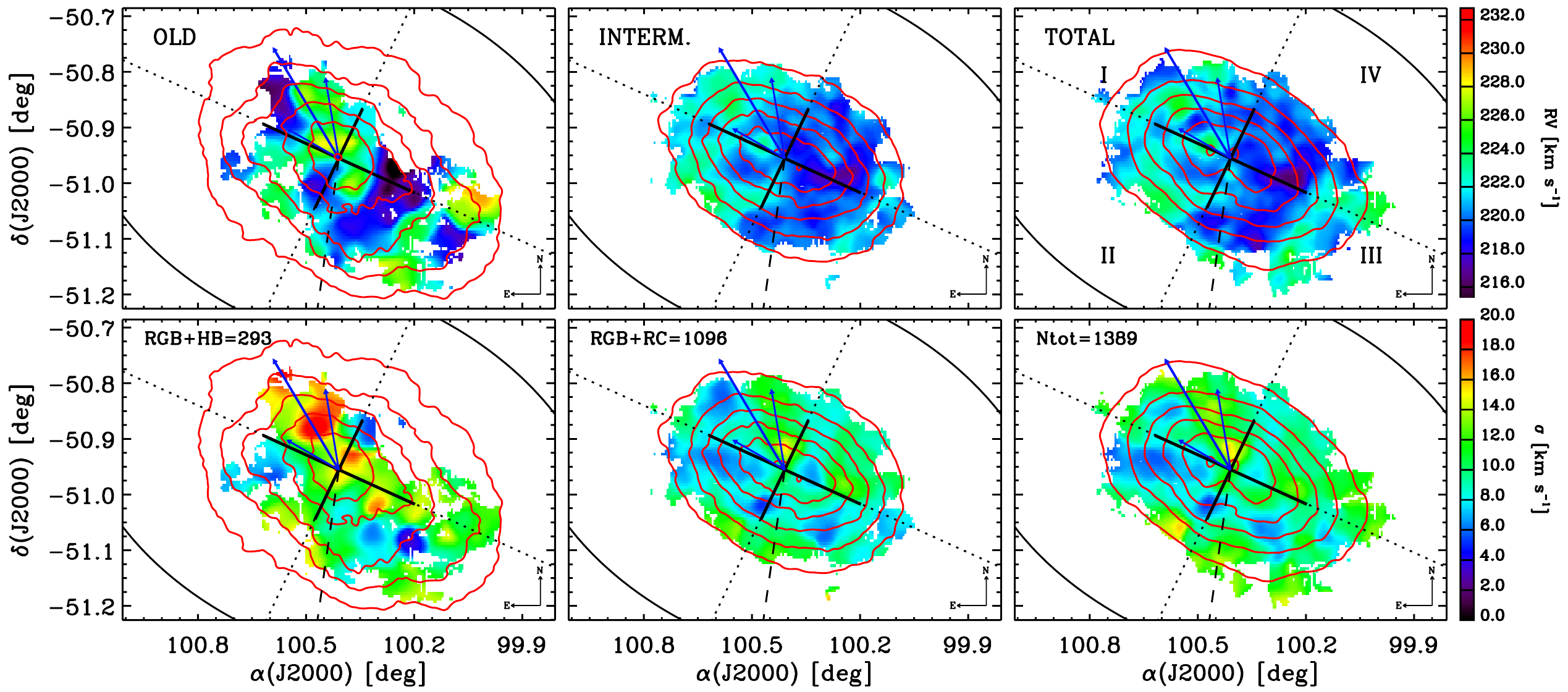}
\caption{Top panels---Left: Radial velocity map of the Carina old stellar 
population (HB, old RGB). 
The RV values were smoothed and color coded. Their range is plotted on the 
right y-axis. The black ellipse shows the Carina tidal radius, while the dotted lines
the major and the minor axis. The extension of solid major and minor axis refers to 
the Carina core radius value \citep{mateo98araa}.
The red iso-contours display radial distribution 
of the old stellar population selected using photometric criteria (see Fig.\ref{contour}). 
The arrows display the Carina proper motion and its uncertainty
according to \citet[][S.~Piatek private communication]{metz08}.
The dashed black line shows the direction of the Galactic center. 
Middle: Same as the left panel, but for the intermediate-age stellar population
(RC, intermediate-age RGB). 
Right: Same as the left panel, but for the entire spectroscopic sample
Bottom panels---Same as the top panels, but for the RV dispersion.      
\label{RVmap}}  
\end{figure*}

The link between the red and the blue RGB stars with the old and the 
intermediate-age stellar population has been discussed in detail in 
\citet{monelli14} and in \citet{fabrizio15}. However, previous considerations 
were mainly based on photometric properties and leading evolutionary arguments 
concerning their distribution along the RGB and the ratio between old and 
intermediate-age Carina stellar populations. 
Here, we take advantage of the new homogeneous and accurate RV measurements,
including a sizable sample of solid tracers of the old stellar population, to 
further validate the \cubi\ index.  

The top panel of Fig.~\ref{hist_rgb} displays the RV distribution of candidate 
intermediate-age RGB stars (blue shaded area), while the bottom panel shows the same 
distribution but for candidate old RGB stars. 
Quantitative constraints based on the $\chi^2$ of 
the two RV distributions indicate that they are different at the 99\%\ confidence level.    
In particular, we found that the old component attains similar (after bi-weight cleaning) 
mean RV ($\mu$$\sim$221~\kms), but larger RV dispersion ($\sigma$=11.7~\kms) 
compared with the intermediate-age one ($\sigma$=9.7~\kms). To further validate the 
above difference, we performed several numerical simulations in which we randomly 
extracted from the intermediate-age RGB sub-population (682 stars) the same number of 
stars (219) included in the old RGB sub-population. We found that the sample size does 
not affects the difference between the two sub-populations, indeed the mean over the 100 
experiments gives a mean RV of 220.6~\kms\ and RV dispersion of 9.6~\kms.
Interestingly enough, we found a similar difference in mean RV and RV dispersion values 
between HB stars (pure old component) and RC stars (pure intermediate-age component). 
The latter component has a mean RV of 219.9~\kms\ and a RV dispersion of 10.2~\kms, 
while the former one has 223.1 and 16.2~\kms, respectively. We performed a series of 
numerical experiments by randomly extracting 74 stars (number of HB stars) from 
the 414 RC stars, and we found very stable values for the mean RV and the 
RV dispersion, namely 220.1 and 10.0~\kms.

The current findings are providing an independent kinematic support for the 
connection between red and blue RGB stars, selected according to the \cubi\ 
index, with old and intermediate-age stellar populations.
In passing, we note that if the \cubi\ index in dwarf galaxies is mainly 
a metallicity indicator instead of an age indicator, the coupling of the 
more metal-poor sub-population with old HB stars and of the more 
metal-rich sub-population with RC stars is still in-line with the 
kinematic analysis we plan to perform. The results of this investigation 
are not affected by the selection criteria adopted to separate the 
quoted sub-populations. 

The above arguments are soundly supported by the clear difference in 
the sky distribution between the old tracers (red RGB, HB, faint SGB; 
2876 stars)  and the intermediate-age tracers (blue RGB, RC, bright SGB; 
3757 stars). The former one (left panel of Fig.~\ref{contour}) 
is characterized by a broad radial 
distribution and by iso-contours showing clear evidence of asymmetries when 
moving toward the outermost galaxy regions. The latter one (middle panel) 
is more centrally concentrated and it shows smoother radial distribution 
as a function of the radial distance. 
The iso-contours of the entire sample of candidate Carina stars (right 
panel) shows several secondary features (double peak, changes in the 
position angle). Thus suggesting a more complex interaction among the 
different Carina sub-populations.    

These findings fully support previous investigations concerning the 
difference between old and intermediate age stellar populations 
in Carina \citep{monelli03} and in dwarf spheroidals \citep{hidalgo13,monelli16}.  

\section{Radial velocity maps}
\label{sec:rvmap}

\begin{figure*}
\centering
\includegraphics[width=0.98\textwidth]{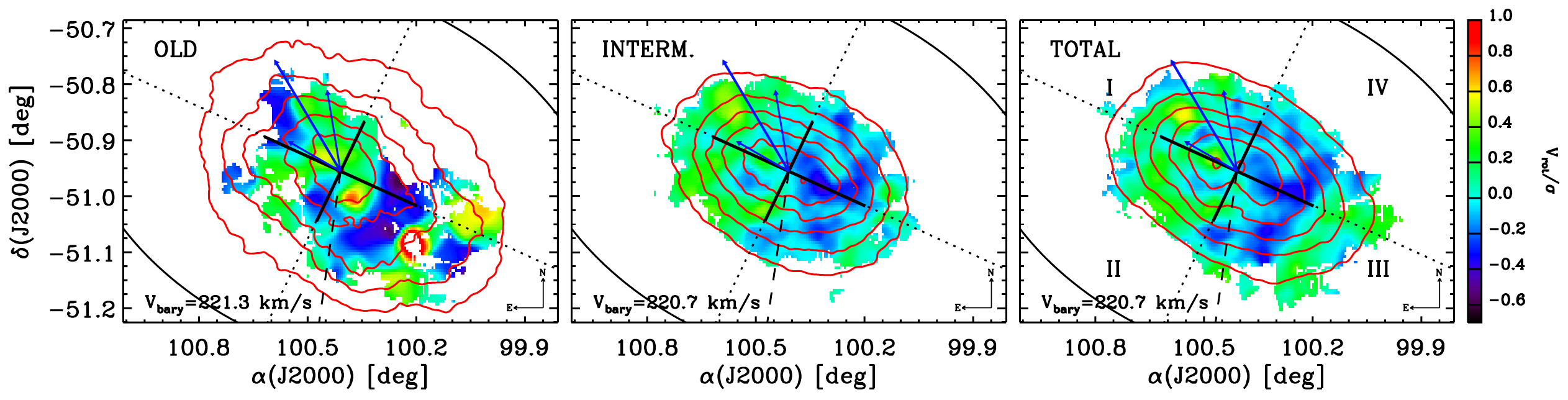}
\caption{Left---RV map of the ratio between $V_{rot}$ and RV dispersion 
for the old stellar population plotted in the right panels of Fig.~\ref{RVmap}. 
The peak of the RV distribution adopted to estimate the $V_{rot}$ is 
also labelled. 
Middle---Same as the left panel, but for the intermediate-age stellar population.  
Right---Same as the left panel, but for the entire spectroscopic sample.  
\label{RVonSigma}}  
\end{figure*}
	
The selection criteria described above allow us, for the first time, 
to handle a sizable sample of kinematic measurements of solid old 
candidate Carina stars (RGB+HB, 293 stars). The top-left panel of 
Figure~\ref{RVmap} 
shows the RV map for the old stellar component. It was computed 
for every star in the sample by averaging over
five nearest objects and smoothing with a Gaussian kernel in 
which the $\sigma$ was assumed equal to the RV dispersion of the 
five nearest stars. The RV map is color coded and the 
RV range is plotted on the right y-axis. To guide the reader, 
we over-plotted the isophotal contours of the old candidate Carina 
stars on top of the RV map (see, Fig.~\ref{contour})

A preliminary analysis of 
the above maps suggests that the old sub-population displays a radial 
velocity distribution with a well defined peak located across the 
innermost regions of the galaxy. The RV map does not show a clear 
pattern, and indeed kinematically hot and cold spots are present in the 
same quadrant. This notwithstanding, there is evidence of a well defined 
increase in radial velocity inside the core radius (see the solid semi-axis). 
Indeed these regions in the I, III and IV quadrant attain RVs that 
are either equal or larger than the peak in the RV distribution of the 
old sub-population (221.3~\kms). 

A similar RV map was also computed for the sub-population of intermediate-age 
candidate Carina stars (RGB+RC, 1096 stars). The middle panel of 
Figure~\ref{RVmap} shows the RV map computed by averaging for every 
star in the sample the 25 nearest objects.
For the intermediate-age sub-population we found a clear evidence of an 
axial rotation. In particular, we found two well defined regions covering 
the western (III and IV quadrants) and the eastern (I and II quadrant) 
side of the galaxy, with a mean radial velocity difference with respect to the 
RV peak of the intermediate-age sub-population (220.7~\kms) of --4~\kms\ 
and +5~\kms, respectively.
We performed a number of tests to constrain the statistical significance of the 
above findings. In particular, we randomly extracted a sub-sample of $\sim$250 
stars from the intermediate-age sub-sample and computed once again the RV 
map. We found that the difference in RV pattern between I+II quadrants and 
III+IV quadrants can be clearly detected in all of them. Note that the number 
of stars that we randomly subtract from the intermediate-age sub-sample is 
similar to the size the old age sub-population. We changed the number of stars 
randomly subtracted from 1/4 to 1/3 of the entire sample and we found once 
again the same RV pattern. 

The top right panel of Figure~\ref{RVmap} shows the RV map of the entire 
(old+intermediate-age) spectroscopic sample. The RV map is quite similar to 
the pattern showed by the intermediate-age sub-population. Thus suggesting that 
the global kinematical properties are mainly driven by this sub-population. 
Note that recent star-formation history of this galaxy clearly indicates that the 
intermediate-age subpopulation includes the 70\%\ of the entire stellar content of 
Carina \citep{small13, deboer14, savino15}.    

The bottom left panel of Fig.~\ref{RVmap} shows the RV dispersion of the 
old sub-population. Data plotted in this panel show three interesting features: 
\textit{(a)} the old sub-population attains RV dispersions that are, as expected,
larger than the RV dispersion of the intermediate-age subpopulation.   
\textit{(b)} The largest RV dispersions are approached in the same regions in which 
the RV map suggested a steady increase in RV distribution. Interestingly
enough, values larger than 15~\kms\ are attained in the I quadrant, in the 
regions located across the direction of the Carina proper motion (long blues 
arrows). However, the current uncertainties on the Carina proper motion do not 
allow us to constrain the above evidence. 
\textit{(c)} There is evidence of a kinematically hot region in the first 
quadrant, for $\alpha$=6\hr41\min54.60\sec\ and 
$\delta$=--50\deg53\arcmin6.9\arcsec, 
and of a cold region in the third quadrant. 
We double checked the radial distribution of the hot and cold region and they 
appears real, since they are a mix of HB and old RGB stars. More firm conclusions 
require larger samples of old tracers. 

The bottom middle panel of Fig.~\ref{RVmap} shows the RV dispersion of the 
intermediate-age sub-population. The RV dispersion for this stellar component 
appears quite smooth and homogenous over the entire galaxy body. The RV 
dispersion of this stellar component is on average 50\%\ smaller than 
the old sub-populations. The RV dispersion of the entire (old+intermediate-age, 
bottom right panel) spectroscopic sample is, once again, quite similar to the  RV 
dispersion of the intermediate-age sub-population.   

To constrain on a more quantitative basis the kinematic properties of the 
two quoted sub-populations, we estimated their residual radial velocity 
by subtracting to the entire RV map the peak of their RV distribution, 
namely the quantity $\rm{RV_{local\:mean}-RV_{Carina}}$. 
In the following we define such a residual radial velocity as 
"rotational velocity". The left panel of Figure~\ref{RVonSigma} shows the 
ratio between the "rotational velocity" map and the RV dispersion map 
plotted in the bottom left panel of  Fig.~\ref{RVmap} for the old, 
sub-population. The range of values attained by the $V_{rot}/\sigma$  
ratio are given on the the right y-axis. They are on average systematically 
smaller than 0.5. Thus suggesting that the old sub-population in Carina 
is pressure supported ($V_{rot}/\sigma$$<$1) instead of rotation supported 
($V_{rot}/\sigma$$>$1). The same outcome applies to the intermediate-age 
population (middle panel) in which the ratio $V_{rot}/\sigma$ attains 
even smaller values. However, the asymmetry between I+II and III+IV 
quadrants is still quite clear. The entire spectroscopic sample (right 
panel) shows, once again, a pattern similar to the intermediate-age 
subpopulation. 

The above evidence indicates that Carina is mainly a pressure supported 
stellar system. However, the occurrence of relic rotational map in 
the intermediate-age sub-population seems to be supported by the radial 
velocity map and by the $V_{rot}/\sigma$ map.    
%

\section{Radial velocity variations}
\label{sec:rvvar}
\subsection{Radial changes}
\label{subsec:radial}
	
\begin{figure}
\centering
\includegraphics[width=0.49\textwidth]{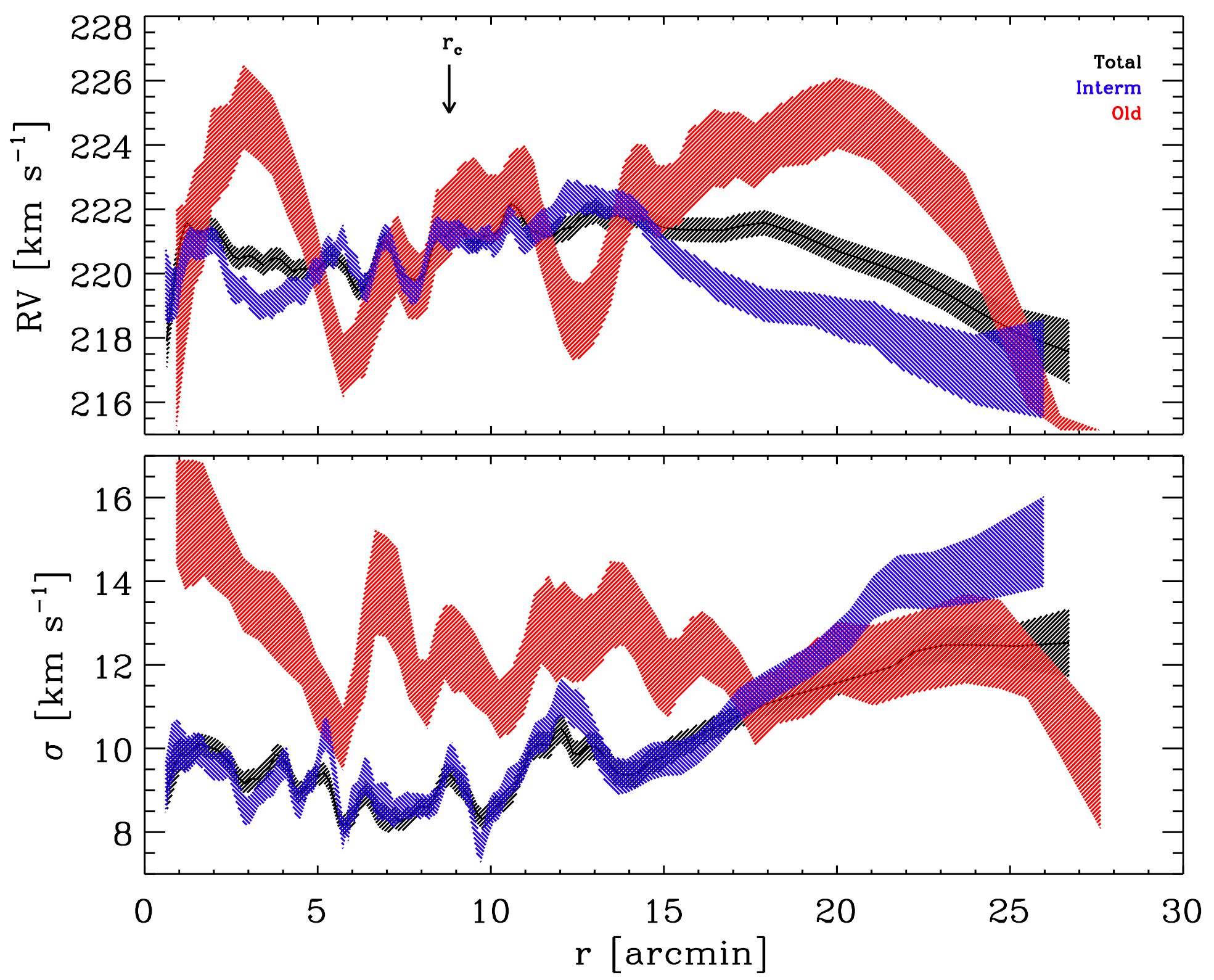}
\caption{
Top---Peak of the RV distribution as a function of the Carina galactocentric distance. 
The red stripe shows the RV for the old stellar population, while the blue one the 
RV for the intermediate-age and the black one the RV for the entire spectroscopic 
sample. The shaded areas take account of the uncertainties in RV distribution    
due to the random subtraction of contaminant field stars and to RV measurements. 
The arrow marks the position of the Carina core radius.
Bottom---Same as the top panel, but for the radial velocity dispersion. 
\label{radial}}  
\end{figure}

To further constrain the kinematics of Carina sub-populations 
we decided to constrain the radial and the angular variations of both 
RV and RV dispersion. The top panel of Fig.~\ref{radial} shows the 
RV variation as a function of the radial Carina distance. Note that the 
radial distance was estimated taking into account the position
angle and the eccentricity (see \citealt{fabrizio11}). 

To avoid spurious fluctuations in the mean RV, we ranked the
entire sample of old Carina stars as a function of the radial distance 
(r [arcmin]) and estimated the running average by using the first
30 objects in the list. The mean distance and the mean velocity of the bin 
were estimated as the mean over the individual distances and RV 
measurements of the 30 objects included in the box.
We estimated the same quantities moving by 5 objects in the 
ranked list until we accounted for the last 5 objects in the 
sample with the largest radial distances. The red shaded area 
shows the RV variation of the old sub-population and takes 
into account the intrinsic errors in the radial velocity measurements and in 
the possible field star contaminants. The approach adopted to 
constrain the latter uncertainty was already discussed in Section~\ref{sec:rvmap}. 

There is evidence that the RV of the old sub-population after an initial 
increase in the innermost galactic regions (r$<$3\arcmin) experiences a decrease 
of almost 8~\kms\ when moving from r$\sim$3\arcmin\ to r$\sim$6\arcmin.
Moreover and even more importantly, the RV dispersion in the same 
regions decreases from $\sim$15~\kms\ to $\sim$10~\kms. Thus suggesting
a strong correlation with the RV variation.  
Estimates available in the literature suggest that the core 
radius of Carina is r$_c$$\sim$8.8\arcmin\ \citep{mateo98araa}, 
however, the above evidence is suggestive of a more compact core region.

The RV of the old stellar component displays, at radial distances larger 
than the quoted secondary minimum, variations of the order of 2~\kms\ up to 
r$\sim$15\arcmin. At larger distances the RV is steadily increasing until 
it approaches a broad secondary maximum ($\sim$225~\kms) and beyond a 
steady decrease until it approaches its absolute minimum ($\sim$215~\kms) in 
the outermost regions of the galaxy. 
The quoted radial variations indicate that the $\sigma$ is almost 
constant over the entire region in which the RV shows the broad maximum. 
The RV and the $\sigma$ approach their absolute minima towards the truncation 
radius of the galaxy (r$_t$$\sim$28.8\arcmin).       

The approach adopted to compute the mean RV of the intermediate-age population 
is the same as of the old one, but the running average was computed including in 
the box 80 stars, and the box was moved in the ranked list of stars using ten
stepping stars. The blue shaded area plotted in Fig.~\ref{radial} shows 
the RV and the RV dispersion of the intermediate-age stellar component. The 
variation is smooth over the entire radial range covered by the current 
spectroscopic data. The variations in RV are on average smaller than 2~\kms, 
while the variations of the RV dispersion are smaller than 1~\kms. However, 
we do have evidence of a steady decrease in RV and of a steady increase in 
RV dispersion for radial distances larger than r$\sim$14\arcmin. Note that 
these are the regions in which the old stellar component attains a broad 
maximum. Moreover, the $\sigma$ of the intermediate-age component for 
r$>$14\arcmin\ displays a mirror trend compared with the RV, since it steadily 
increases and attain its maximum value ($\sim$15~\kms) in the outermost 
galactic regions. The black shaded area display a similar trend, but in dealing 
with the entire sample we included 100 objects in the box and assumed 30 
stepping stars. This is the reason why the upper limit in radial distance is 
between the old- and the intermediate-age stellar component.  

\subsection{Angular changes}
\label{subsec:angul}

\begin{figure}
\centering
\includegraphics[width=0.49\textwidth]{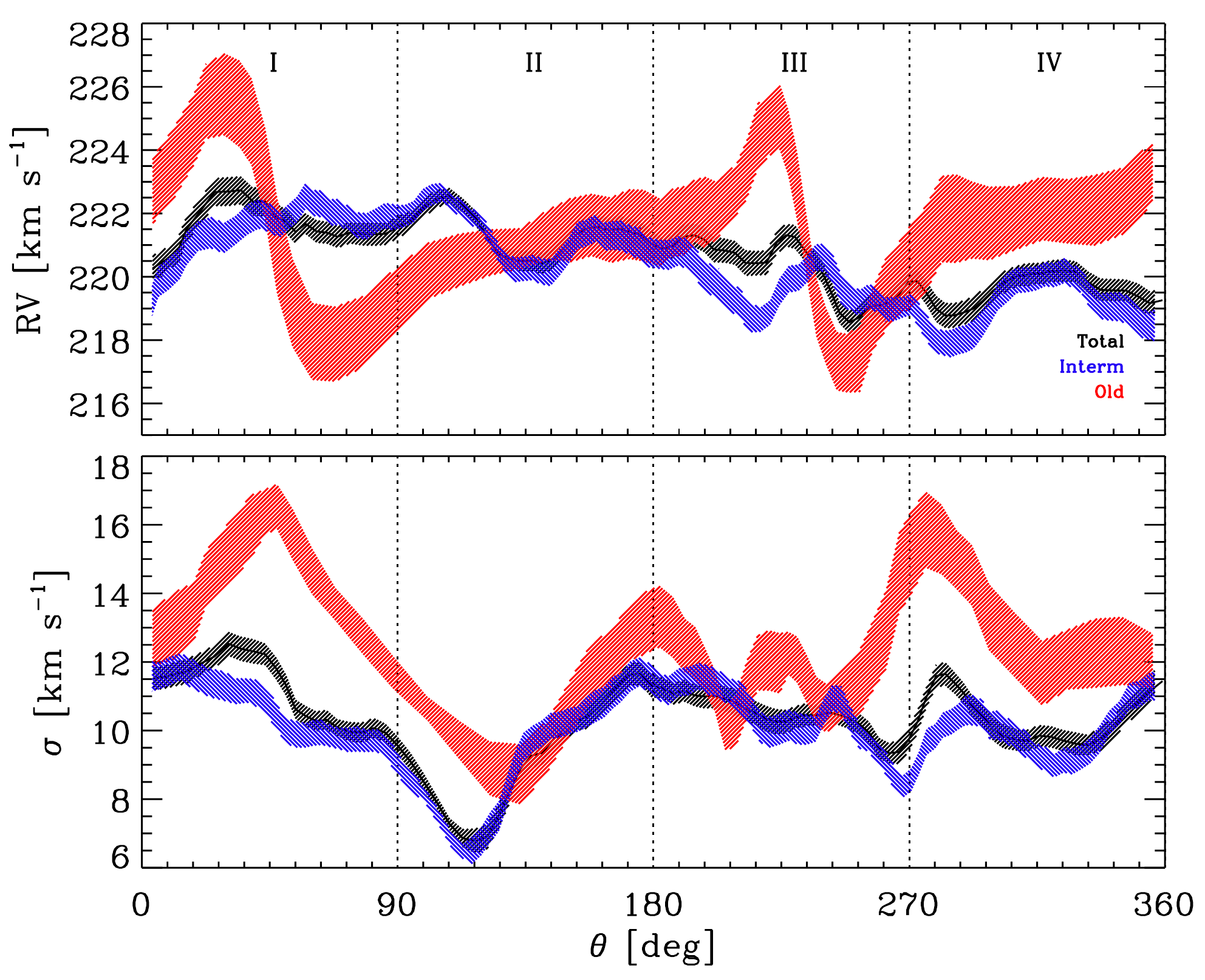}
\caption{Same as Fig.~\ref{radial}, but as a function of the angle 
across the Carina body.\label{angular}}  
\end{figure}

To further constrain the spatial variation in RV and in RV dispersion we 
also computed the running average as a function of the Carina angle. The 
main difference with the approach discussed in Sect.~\ref{subsec:radial} 
is that we use a wedge instead of a box. We ranked the stars from the 
northern minor axis (0\deg) towards the eastern major axis (90\deg) and 
used the usual running average.
For clearness, the quadrants were labelled as in Fig.\ref{contour}.
The angular variations of both old- and intermediate-age sub-populations 
were estimated including the same numbers of stars and the same number of 
stepping stars adopted for the radial variations (see Sect.~\ref{subsec:radial}). 

The shaded areas plotted in Fig.~\ref{angular} follow the same color code 
as in Fig.~\ref{radial}.  Data plotted in the top panel of Fig.~\ref{angular} 
display even more clearly that the old stellar component (red area) experiences 
significant changes in RV inside the I and the III quadrant. The RV in the 
I quadrant moves from its absolute maximum to a secondary minimum and the 
variation is of the order of 8~\kms, while in the III quadrant it moves 
from a secondary maximum to its absolute minimum and the variation is quite 
similar. On the other hand, the RV in the II and in the IV quadrant shows a 
steady and smooth increase of a few \kms. 
The RV dispersion of the old sub-population is also quite interesting, 
since it attains its absolute maximum and minimum ($\Delta\sigma\sim 7$~\kms) 
across the sharp variation in RV of the I quadrant (50\deg$<$$\theta$$<$140\deg). 
On the other hand, the $\sigma$ variation across the sharp change in RV of 
the III quadrant (200\deg$<$$\theta$$<$250\deg) is almost a factor of two smaller. 

The intermediate-age stellar component (blue area) shows smooth variations across 
the four quadrants. There is evidence of a steady decrease of RV when moving 
from the I to the IV quadrant. The global variation is of the order of 
5~\kms, with the absolute maximum located in the I quadrant and the 
absolute minimum in the IV quadrant. The RV dispersion shows a different 
trend, and indeed its entire excursion takes place inside the first two
quadrants, where it changes from $\sim$12~\kms\ at $\theta$$\sim$0\deg\
to $\sim$6~\kms\ at $\theta$$\sim$120\deg.   

The above empirical evidence are suggesting that the sharp changes in RV 
and in $\sigma$ of the old sub-population are taking place along the 
direction of the Carina proper motion. Once again the intermediate-age 
population appears reminiscent of a rotation pattern and the variations 
are smooth across the body of the galaxy.  
 
\section{Comparison with $N$-body simulations}
\label{sec:simul}

\begin{figure}
\centering
\includegraphics[width=0.45\textwidth]{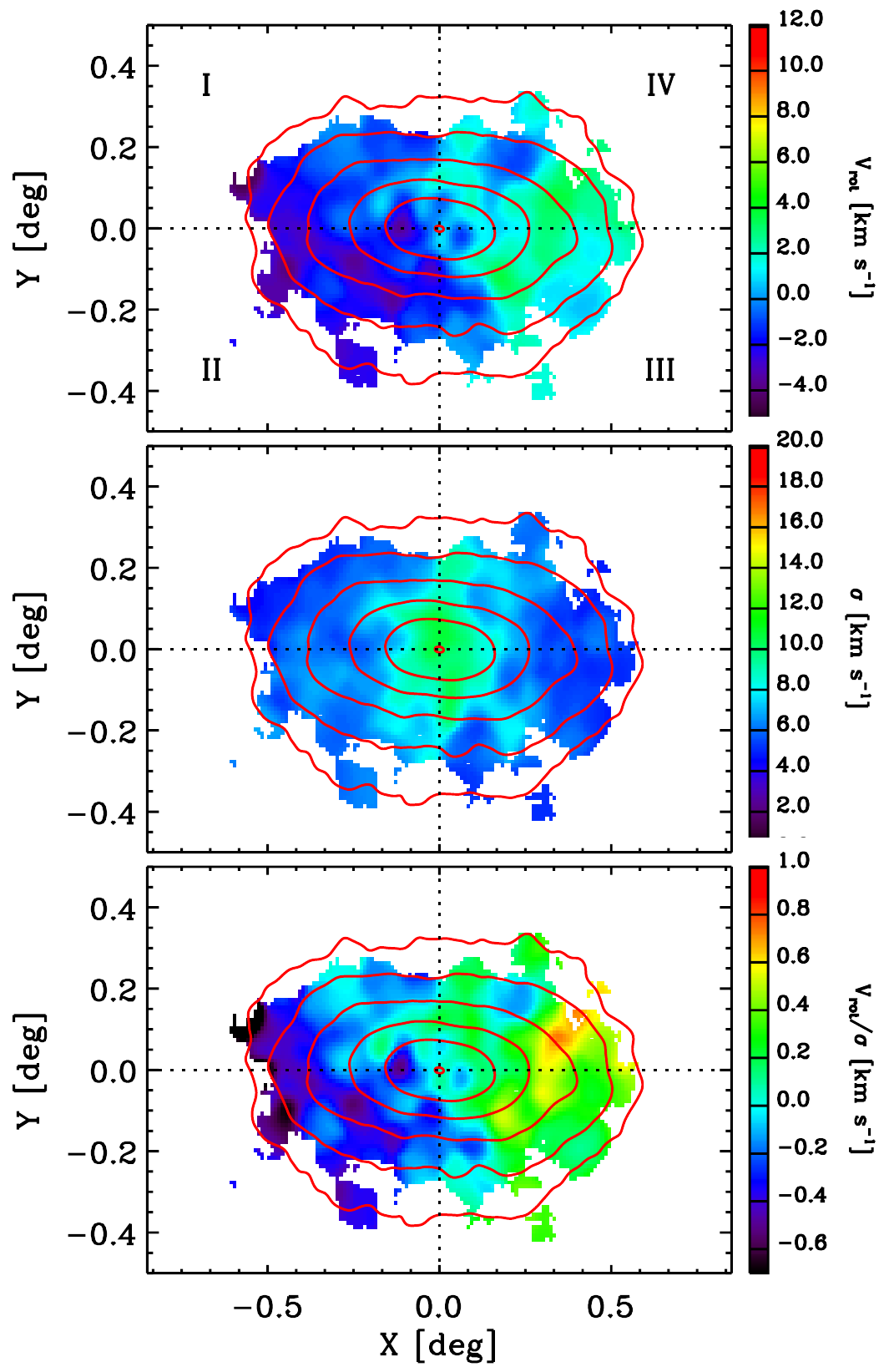}
\caption{Top---Rotational velocity map based on the 
{\it N}-body model (view angle of 60\deg\ with respect to the major axis). 
The criteria adopted to select 
predicted velocities are discussed in detail in the text.  
Middle---Same as the top panel, but for the velocity dispersion. 
Bottom---Same as the top panel, but for the $V_{rot}/\sigma$ ratio.
\label{simulmap}}  
\end{figure}

In this section we present a preliminary comparison of the observed radial 
velocity distributions with $N$-body simulations described in \citet{lokas15}. 
The velocity and dispersion maps of Fig.~\ref{RVmap} and the stellar density 
distributions of Fig.~\ref{contour} show very similar characteristics to the 
dwarf spheroidal galaxies formed as a result of tidal interactions of an 
initially disky dwarf galaxy orbiting a Milky Way-like host (e.g., see Fig.~5 in 
\citealt{lokas15} and Fig.~\ref{simulmap} in \citealt{ebrova15}). In this simulation 
(I0) the dwarf galaxy disk had an exactly prograde orientation with respect 
to the orbit and the galaxy underwent very strong transformation. 
At the first (out of five) pericenter passage the disk evolved into a bar 
which during the subsequent evolution became shorter and thicker. For the 
comparison with the data for Carina we used the simulation output saved after 
8.5~Gyr from the start of the evolution, corresponding to the time after the 
fourth pericenter passage when the dwarf is at the apocenter of the orbit. 
At this time, most of the dwarf's initial rotation was already replaced by 
random motions and the simulated dwarf galaxy is approximately in equilibrium 
after being perturbed at the previous pericenter passage.

%

\begin{figure}
\centering
\includegraphics[width=0.45\textwidth]{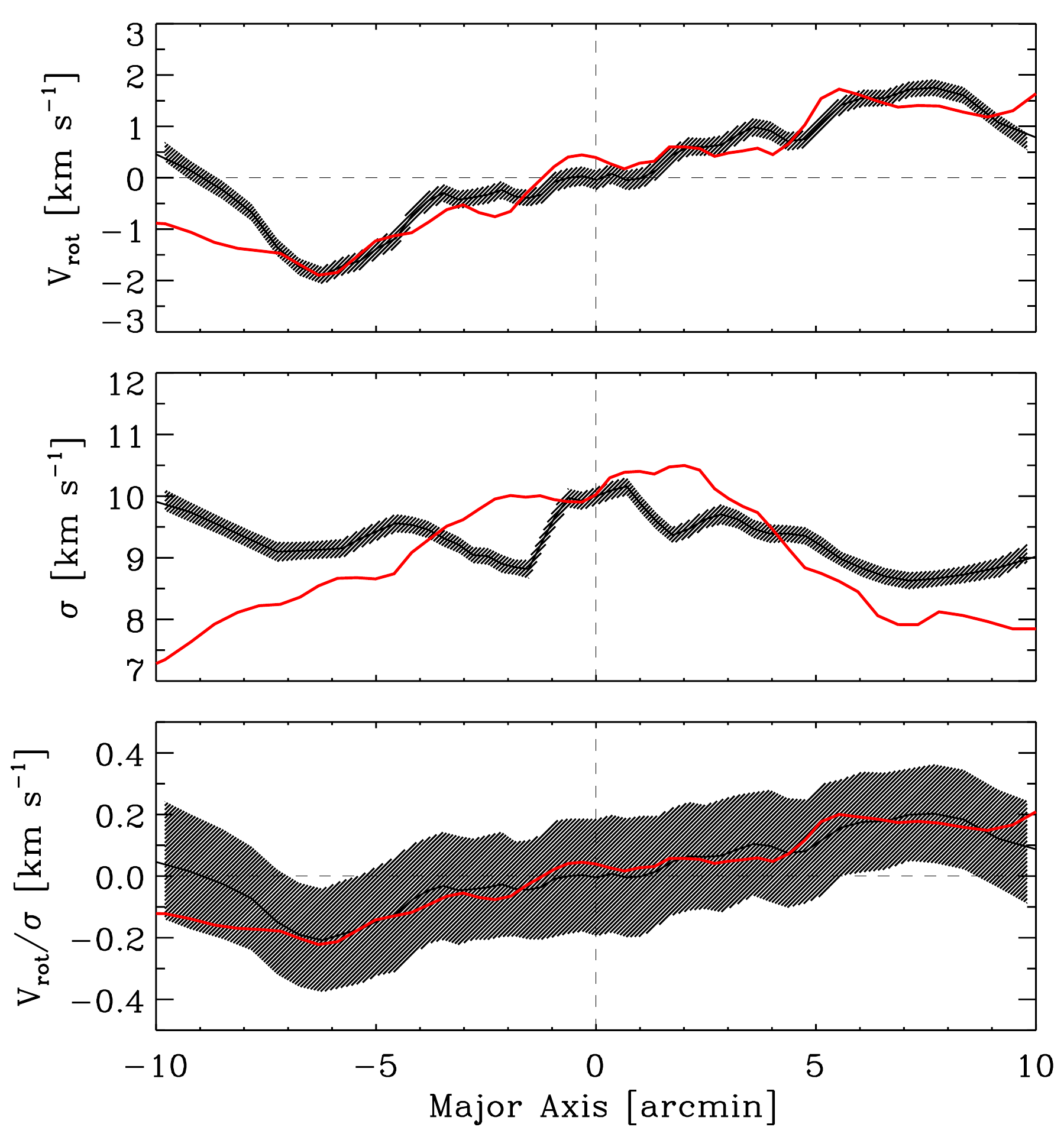}
\caption{Top---Comparison between predicted (red line) and observed 
(black line) rotational velocity measured along the major axis of the 
Carina central regions.
Middle---Same as the top panel, but for the velocity dispersion.
Bottom---Same as the top panel, but for the $V_{rot}/\sigma$ ratio.
\label{simulprofiles}}  
\end{figure}

To properly compare theory and observations, the simulations were rescaled to 
the Carina distance ($D_\sun$=105$\pm$6~kpc, \citealt{pietrzynski09}). Moreover, 
the Carina barycentric velocity (220.7~\kms) was also subtracted from the 
observed RV distribution. We investigated different lines of sight and the 
agreement between theory and observations was found to be best when the 
simulated dwarf galaxy was observed at an angle of 60\deg\ with respect to 
the major axis of the stellar component, but keeping the observer on the 
orbital plane of the dwarf.

We constructed the radial velocity maps of the simulated dwarf following 
the same approach adopted for measuring the RV distribution 
(Fig.~\ref{RVmap}) and the RV dispersion (Fig.~\ref{RVonSigma}) of the 
observations. We randomly selected from the simulations a sub-sample of 
2000 stars, i.e., a number of objects similar to the observed sample. 
In order to account for instrumental errors, the predicted RV measurements
were convolved with a Gaussian distribution with a $\sigma$ of 4~\kms. 
Figure~\ref{simulmap} shows, 
from top to bottom, the maps for the rotational velocity, the velocity 
dispersion and the $V_{rot}/\sigma$ ratio for the selected simulation 
output. On top of the above maps we also plotted the surface density contours 
calculated from a larger sample of 36,000 stars, corresponding to the 
number currently available in photometric studies. 
Figure~\ref{simulprofiles} shows the comparison between the rotational 
velocity, the velocity dispersion and the $V_{rot}/\sigma$ ratio 
measured in the core along the major axis of both the simulated 
and the real dwarf.

Given the fact that the simulation discussed here was not performed to 
reproduce the observed properties of Carina, the similarity of the mock 
velocity maps and profiles to the real data for Carina is astonishing. 
The comparison strongly argues for a formation scenario in which Carina 
started as a disky dwarf and experienced strong, possibly multiple, tidal 
interactions with the Milky Way. As a result, it transformed from a disk 
to a prolate spheroid and its rotation was replaced by random motions. 
However, some residual rotation around the minor axis remained as a relic 
of its initial conditions. For this process to be efficient, Carina must 
have been accreted on a prograde, rather than a retrograde orbit \citep{lokas15}. 
We performed several numerical experiments using $N$-body simulations with 
different initial conditions (prograde, retrograde orientations) and different 
epochs after the starting of their evolution. In some cases we found a good 
match either of the shape (I90) or of the $V_{rot}/\sigma$ ratio (I270). 
However, the best fit of both the shape and the kinematics is only obtained 
for the above prograde orbit.

\section{Summary and final remarks}
\label{sec:summary}

We present new radial velocity measurements of Carina evolved 
stars using optical low-resolution spectra (R$\sim$600) collected with 
VIMOS a multi-object slit spectrograph available at the VLT. 
The spectra were collected using the MR-GG475 grism, with a 
wavelength coverage of 4,800$\div$10,000\AA. We secured, with 
$\sim$15~hr of exposure time, 2454 individual spectra for 
332 stars. The targets have visual magnitudes ranging from 
\vv=20 to \vv=21.5~mag and they are either old HB stars 
or intermediate-age RC stars or RGB stars. The new data 
were complemented with archival optical medium- and high-resolution 
spectra collected with FORS2 (R$\sim$2,500), GIRAFFE-LR 
(R$\sim$6,500), GIRAFFE-HR (R$\sim$20,000) and UVES 
(R$\sim$40,000) spectrographs at VLT. The spectroscopic 
data were collected on a time interval of 12 years.  
We ended up with a sample of more than 21,000 spectra for 2112 stars. 

The above data were complemented with radial velocity measurements 
for more than 1500 stars available in the literature and based on 
high-resolution spectra \citep{walker07}. As a whole we have radial velocity 
measurements for 2748 stars and among them 1389 are candidate 
Carina stars. 
The current spectroscopic dataset covers 
the entire body of the galaxy, with the high-resolution spectra that 
are more centrally concentrated, while medium- and low-resolution 
spectra approach the truncation radius.     

Special attention was paid in the validation of radial velocity 
measurements among the different spectroscopic samples. Moreover, 
the sizable sample of targets in common allowed us to perform an 
accurate calibration of radial velocities based on low-resolution 
spectra. 

We also took advantage of the superb multi-band (\uu\bb\vv\ii) photometric 
catalog that our team collected during the last twenty years, and in 
particular, of the \cubi\ index to select candidate old and 
intermediate-age sub-populations. 

Interestingly enough, we found that the radial velocity distributions of "cooler" and 
"hotter" (according to the \cubi\ index) RGB stars attain dissimilar values for (biweight) 
mean RV and dispersion and they are different at the 99\%\ confidence level. On the 
other hand, the distribution of "hotter" RGB stars shows a standard deviation similar to 
RC stars, that are intermediate-age tracers, distribution. The same accounts between 
the distributions of "cooler" RGB and HB stars, that are old stellar tracers.
This result provides an independent kinematic support 
to the use of \cubi\ as either an age and/or a metallicity indicator.

The photometric and kinematic selections allowed us to build a sample 
of $\sim$300 old (HB+RGB) and of $\sim$1100 intermediate-age (RC+RGB)  
stellar tracers. The RV distribution of the old sub-population attains 
peaks inside and outside the core radius that are slightly larger than 
for the intermediate-age sub-population, but the difference is smaller 
than 1$\sigma$. However, the standard deviation of the former sample 
is systematically larger (2-3~\kms) than the latter one.   
The RV map of the old sub-population does not show evidence 
of a pattern, it is characterized by an unevenly distribution of hot 
and cold spots across the body of the galaxy. The RV dispersion attains
values larger than 15~\kms\ in the direction of the Carina proper motion.  
However, the current uncertainties on proper motion do not allow us to 
constrain this empirical evidence. 

The RV map of the intermediate-age sub-population shows evidence of an 
axial rotation between the western (III and IV quadrant) and the 
eastern (I and II quadrant) side of the galaxy. The difference in 
RV between the main peak and the above regions is of --4 and +5~\kms. 
The RV dispersion of the intermediate-age sub-populations has a smooth 
variation across the entire body of the galaxy. Inside the core radius 
it is roughly 50\%\ smaller than the RV dispersion of the old stellar 
component. 

The RV maps were also adopted to estimate the rotation velocity, i.e., 
the difference between the local value of the RV and the main peak of the 
Carina RV distribution. We found that both the old and the intermediate-age 
sub-populations attain values of the $V_{rot}/\sigma$ ratio that are 
systematically smaller than one, thus suggesting that Carina is a pressure 
supported instead of a rotation supported stellar system.    
   
The kinematics of the old sub-population shows a complex variation as a function 
of the radial distance. The RV and the RV dispersion display their entire 
excursion inside the core radius. The same parameters show changes of the order 
of 2~\kms\ in the outer most regions and attain their absolute minima across 
the truncation radius. On the other hand, the intermediate-age sub-population 
shows smooth radial variations, but for radial distances larger than 
15\arcmin\ the RV is steadily decreasing by more than 3~\kms\ while the 
RV dispersion is steadily increasing by more than 4~\kms.   

The old stellar component also shows a complex angular variation. It 
experiences significant variations both in the I and in III quadrant 
with RV that changes inside the above gradients by $\sim$8~\kms. 
The RV dispersion shows similar variations in the I quadrant, but they 
are significantly smaller in the III one. Firm conclusions concerning the 
difference do require larger samples of the old sub-population.  
The intermediate-age stellar component shows, once again, smooth angular 
variations. It is worth mentioning that the RV dispersion of this sub-population 
shows a trend similar to the old sub-populations in the first two quadrants. 
However, both the absolute maximum and the absolute minimum do not appear to be 
symmetric, and indeed the extrema of the former component anticipate the 
latter ones.     

In order to provide more solid constraints on kinematical structure of Carina, 
we performed a detailed comparison between observed and predicted RV distributions 
based on $N$-body simulations for a former disky dwarf galaxy orbiting a giant 
Milky Way-like galaxy recently provided by \citet{lokas15}. 
In the comparison with Carina, we selected a simulation in which the dwarf 
galaxy disk had a prograde orientation with respect to its orbit and 8.5~Gyr
after the start of its evolution, i.e., after its fourth pericenter passage.    
We found that the best fit between theory and observations is attained when
the simulated galaxy is viewed at an angle of 60 degrees 
with respect to its major axis. 

The values attained by the predicted $V_{rot}/\sigma$ ratio are in remarkable 
agreement with observations over the central regions of the galaxy 
(--10\arcmin$\lesssim$r$\lesssim$10\arcmin). This finding becomes even 
more compelling if we take account for the fact that the adopted theoretical 
framework is not specific for Carina. Moreover and even more importantly, it 
is strongly suggesting that the current Carina dSph was a disky dwarf galaxy 
that underwent several strong tidal interactions with the Milky Way. These 
interactions transformed Carina from a disky to a prolate spheroid and its 
rotational velocity was steadily transformed into random motions.  
The rotational velocity we detected along the major axis is still a relic 
of its initial conditions. 

Our findings make Carina an excellent example of the process of the 
formation of dwarf spheroidal galaxies via tidal stirring of disky 
dwarfs. This argument is further supported by the fact that recent 
estimates indicate that the Carina kinematics is strongly dominated 
by its dark matter halo. Indeed, \citet{lokas09}, using the 
radial velocity distribution provided by \citet{walker09a} and 
assuming that mass follows light, found a mass-to-luminosity 
ratio M/L=66$\pm$31 (solar units). 
The above hypothesis that mass follows light is quite plausible for 
strongly tidally stirred stellar systems like Carina, since the dark 
matter halo is trimmed down and it does not extend well beyond the 
stellar component. 

The increase in the sample size and in the spatial sampling is suggesting 
an increase in the RV dispersion. This would imply a steady increase in 
the M/L ratio. However, current findings are suggesting that the RV dispersion 
of the old stellar component is systematically larger than the RV dispersion 
of the intermediate-age component. The latter sample account for at least two 
third of the Carina stellar content, thus suggesting that a detailed knowledge 
of the kinematics of the different stellar components is mandatory to constrain 
the M/L ratio of nearby stellar systems. The increase in sample size and in 
spatial coverage implies a steady decrease in the sampling error associated
with the M/L ratio.   

In a forthcoming investigation, aimed at specifically reproducing 
Carina, we will use $N$-body simulations including gas dynamics and 
star formation to model different stellar populations, as well as the 
known constraints on the Carina's orbit to study the possible 
contamination from the tidal tails \citep{klimentowski07,klimentowski09}. 
This is an ambitious project, 
because it requires a substantial improvement in the sample size of the 
old stellar component and in the galaxy regions located across and 
beyond the truncation radius. The observing and data reduction 
strategy we devised with VIMOS appears a very good viaticum. It 
goes without saying that the use of the ground layer adaptive optics 
facility at UT4 of VLT (GRAAL, \citealt{siebenmorgen11}) and the new integral field 
spectrograph MUSE \citep{bacon06} will also open new paths in constraining the 
kinematic properties of nearby dwarf galaxies.

\begin{acknowledgements}
It is a pleasure to acknowledge the anonymous referee
for her/his comments and suggestions that improved the content and 
the readability of our manuscript.
M.F. acknowledges financial support from the PO FSE Abruzzo 2007-2013
through the grant "Spectro-photometric characterization of stellar populations
in Local Group dwarf galaxies" prot.89/2014/OACTe/D (PI:~S.~Cassisi). 
G.B. thanks the Japan Society for the Promotion of Science
for a research grant (L15518).
M.N. acknowledges support from PRIN-INAF 2014 1.05.01.94.02.
E.L.{\L}. acknowledges support of the 
Polish National Science Center under grant 2013/10/A/ST9/00023.
\end{acknowledgements}

\bibliography{biblio}

\end{document}

%% file: defs.tex
\def\deg{${}^\circ$}
\def\arcmin{${}^{\prime}$}
\def\arcsec{${}^{\prime\prime}$}
\def\hr{${}^{h}$}
\def\min{${}^{m}$}
\def\sec{${}^{s}$}

\newcommand{\giraffe}{{FLAMES\slash GIRAFFE}}

\newcommand{\caii}{\hbox{Ca\,{\scriptsize II}}}

\newcommand{\teff}{\hbox{T$_{\mathrm{eff}}$}}
\newcommand{\logg}{\hbox{$\log g$}}
\newcommand{\feh}{\hbox{[Fe/H]}}

\newcommand{\ms}{\hbox{m\,s$^{-1}$}}
\newcommand{\kms}{\hbox{km\,s$^{-1}$}}

\newcommand{\afe}{\hbox{[$\alpha$/Fe]}}

\newcommand{\uu}{\hbox{\it U\/}}
\newcommand{\bb}{\hbox{\it B\/}}
\newcommand{\vv}{\hbox{\it V\/}}

\newcommand{\ii}{\hbox{\it I\/}}

\newcommand{\umb}{\hbox{\uu--\bb\/}}

\newcommand{\bmv}{\hbox{\bb--\vv\/}}
\newcommand{\bmi}{\hbox{\bb--\ii\/}}

\newcommand{\cubi}{\hbox{{\it c}$_{\rm U,B,I}$\/}}

%% file: tab_logobs.tex
\begin{deluxetable*}{cccccccc}
\tabletypesize{\footnotesize}
\tablewidth{0pt}
\tablecaption{Log of VIMOS observations\label{tab:logobs}}

\tablehead{
Date &  
Pointing &  
$\alpha$(J2000) & 
$\delta$(J2000) & 
Exp.Time & 
Seeing\tablenotemark{a} & 
Airmass & Num. Slits \\
        &           & hh:mm:ss.ss      & dd:mm:ss.s         &       s           &     \arcsec\          &   & }
\startdata
07 Dec 2012 & Deep\_North\_1 &  06:42:27.93 & --50:56:25.5 & 2640 &  0.73-0.60  & 1.122 & 152 \\ 
23 Dec 2012 & Deep\_North\_2 &  06:42:27.93 & --50:56:25.5 & 2640 &  0.75-0.65  & 1.166 & 152 \\ 
08 Dec 2012 & Deep\_North\_3 &  06:42:27.93 & --50:56:25.5 & 2640 &  0.60-0.90  & 1.126 & 152 \\ 
09 Dec 2012 & Deep\_North\_4 &  06:42:27.93 & --50:56:25.5 & 2640 &  1.12-0.73  & 1.127 & 152 \\ 
09 Dec 2012 & Deep\_North\_5 &  06:42:27.93 & --50:56:25.5 & 2640 &  0.83-0.66  & 1.120 & 152 \\ 
23 Dec 2012 & Deep\_North\_6 &  06:42:27.93 & --50:56:25.5 & 2640 &  0.54-0.60  & 1.118 & 152 \\ 
07 Dec 2012 & Deep\_North\_7 &  06:42:27.93 & --50:56:25.5 & 2640 &  0.75-1.13  & 1.125 & 152 \\ 
\\  
31 Dec 2012 & Deep\_South\_1 &  06:40:56.00 & --51:03:49.9 & 2640 &  0.75-0.67  & 1.174 & 161 \\ 
01 Jan 2013 & Deep\_South\_2 &  06:40:56.00 & --51:03:49.9 & 2640 &  0.98-1.11  & 1.117 & 161 \\ 
01 Jan 2013 & Deep\_South\_3 &  06:40:56.00 & --51:03:49.9 & 2640 &  1.20-1.38  & 1.153 & 161 \\ 
01 Jan 2013 & Deep\_South\_4 &  06:40:56.00 & --51:03:49.9 & 2640 &  1.00-1.20  & 1.143 & 161 \\ 
02 Jan 2013 & Deep\_South\_5 &  06:40:56.00 & --51:03:49.9 & 2640 &  0.75-0.82  & 1.250 & 161 \\ 
02 Jan 2013 & Deep\_South\_6 &  06:40:56.00 & --51:03:49.9 & 2640 &  1.31-0.84  & 1.149 & 161 \\ 
02 Jan 2013 & Deep\_South\_7 &  06:40:56.00 & --51:03:49.9 & 2640 &  0.76-0.66  & 1.117 & 161 \\ 
\\ 
03 Jan 2013 & Shallow\_North\_1 &  06:42:27.93 & --50:56:25.5 & 2640 &  0.90-1.11  & 1.121 & 107 \\ 
03 Jan 2013 & Shallow\_North\_2 &  06:42:27.93 & --50:56:25.5 & 2640 &  1.16-0.84  & 1.176 & 107 \\ 
04 Jan 2013 & Shallow\_North\_3 &  06:42:27.93 & --50:56:25.5 & 2640 &  0.93-0.96  & 1.116 & 107 \\ 
\\ 
06 Jan 2013 & Shallow\_South\_1 &  06:40:56.00 & --51:03:49.9 & 2640 &  0.69-0.85  & 1.131 & 110 \\ 
06 Jan 2013 & Shallow\_South\_2 &  06:40:56.00 & --51:03:49.9 & 2640 &  0.90-0.99  & 1.119 & 110 \\ 
06 Jan 2013 & Shallow\_South\_3 &  06:40:56.00 & --51:03:49.9 & 2640 &  0.84-0.82  & 1.167 & 110 \\ 
\enddata
\tablenotetext{a}{Seeing condition at the begin and the end of exposure.}
\end{deluxetable*}

%% file: tab_RV.tex
\begin{deluxetable*}{lccccccrr}
\tabletypesize{\footnotesize}
\tablewidth{0pt}
\tablecaption{Radial velocities of spectroscopic targets.\label{tab:RV}}
\tablehead{
ID &  $\alpha$(J2000) & $\delta$(J2000) & \vv & \bb & \ii & RV & Spect.\tablenotemark{a} & Pop. Flag\tablenotemark{b} \\        
   & deg & deg & mag & mag & mag & \kms & & }
\startdata
    107664 &       100.4775 &       --50.9502 &    17.685 $\pm$    0.003 &    19.147 $\pm$     0.001 &    16.098 $\pm$    0.003 &     230.1 $\pm$    0.1 &          1,3,4 &    1\\
    111206 &       100.4994 &       --51.0315 &    17.702 $\pm$    0.001 &    19.087 $\pm$     0.001 &    16.146 $\pm$    0.003 &     211.5 $\pm$    1.5 &            3,6 &    1\\
     32411 &        99.98250 &       --50.9602 &    17.883 $\pm$    0.001 &    19.048 $\pm$     0.004 &    16.464 $\pm$    0.002 &     233.7 $\pm$    0.1 &              1 &    1\\
     78252 &       100.3008 &       --51.2212 &    17.896 $\pm$    0.001 &    19.119 $\pm$     0.001 &    16.518 $\pm$    0.004 &     234.4 $\pm$    1.2 &            3,6 &    1\\
    102069 &       100.4432 &       --51.0230 &    17.906 $\pm$    0.001 &    19.152 $\pm$     0.001 &    16.497 $\pm$    0.001 &     213.0 $\pm$    7.3 &            1,3 &    1\\
    107139 &       100.4744 &       --50.9697 &    17.907 $\pm$    0.001 &    19.100 $\pm$     0.002 &    16.521 $\pm$    0.001 &     216.8 $\pm$    7.3 &              1 &    1\\
\enddata
\tablecomments{Table~\ref{tab:RV} is published in its entirety in the electronic edition of the Astronomical Journal. A portion is shown here for guidance regarding its form and
content.}
\tablenotetext{a}{Adopted spectrographs: 1=UVES; 2=Gir-HR; 3=Gir-LR; 4=FORS2; 5=VIMOS; 6=MMFS}
\tablenotetext{b}{Stellar population flags: 0=Field candidates; 1=Old RGB candidates; 2=Intermediate RGB candidates; 3=Horizontal Branch; 4=Red Clump}
\end{deluxetable*}

%% file: tab_gauss.tex
\begin{deluxetable}{cccc}
\tabletypesize{\footnotesize}
\tablewidth{0pt}
\tablecaption{The parameters of multi-Gaussian fit of the RV distribution 
in the form: $y=A_0 \exp{\left[-\frac{(x-\mu)^2}{2\sigma^2}\right]}$.\label{tab:multigauss}}

\tablehead{Component          & $A_0$ & $\mu$ [\kms] & $\sigma$ [\kms]}
\startdata
Field1\tablenotemark{a}            &     9.92$\pm$    0.02 &    32.12$\pm$    0.28 &    22.92$\pm$    0.21 \\
Field2\tablenotemark{a}            &     5.15$\pm$    0.08 &    66.76$\pm$    0.36 &    11.28$\pm$    0.32 \\
Field3\tablenotemark{a}            &     4.22$\pm$    0.02 &    96.86$\pm$    1.01 &    15.38$\pm$    0.63 \\
Field4\tablenotemark{a}            &     2.55$\pm$    0.01 &   145.09$\pm$   13.82 &    88.44$\pm$    6.45 \\
Carina\_B\tablenotemark{b}         &    20.00$\pm$    0.16 &   219.93$\pm$    0.02 &    14.58$\pm$    0.03 \\
Carina\_N\tablenotemark{b}         &    28.41$\pm$    0.22 &   221.67$\pm$    0.01 &     7.83$\pm$    0.01 \\ \hline
Carina\_B--clean\tablenotemark{c}  &    16.05$\pm$    0.22 &   219.55$\pm$    0.05 &    15.98$\pm$    0.08 \\
Carina\_N--clean\tablenotemark{c}  &    31.99$\pm$    0.30 &   221.58$\pm$    0.01 &     8.21$\pm$    0.01 
\enddata

\tablenotetext{a}{Gaussians adopted to field the RV distribution of candidate field stars.}
\tablenotetext{b}{Broad and narrow Gaussians adopted to fit the RV distribution of candidate Carina stars.}
\tablenotetext{c}{Broad and narrow Gaussians adopted to fit the RV distribution of candidate Carina stars after the random cleaning of candidate field stars.}
\end{deluxetable}

%% file: tab_mean.tex
\begin{deluxetable*}{cccc|ccc|ccc}
\tabletypesize{\footnotesize}
\tablewidth{0pt}
\tablecaption{The biweighted mean radial velocity (with error), dispersion and number of stars values are listed for the total sample, intermediate-age and old population. The rows refer to the entire spatial extension of spectroscopic targets, inside and outside the core radius \citep[$r_c=8.8'\pm1.2$\arcmin,][]{mateo98araa}.\label{tab:mean}}
\tablehead{
Radius &  \multicolumn{3}{c|}{Total} &  \multicolumn{3}{c|}{Interm.} &  \multicolumn{3}{c}{Old}\\        
           &    RV [\kms] & $\sigma$ [\kms]& N &  RV [\kms]& $\sigma$ [\kms]& N & RV [\kms]& $\sigma$ [\kms]& N}

\startdata
ALL &220.73$\pm$0.28 & 10.51 &  1389 & 220.66$\pm$0.30 &  9.88 &  1096 & 221.33$\pm$0.75 & 12.91 &   293\\
$r<r_c$ &220.37$\pm$0.36 &  9.91 &   748 & 220.29$\pm$0.37 &  9.34 &   636 & 221.04$\pm$1.27 & 13.42 &   112\\
$r>r_c$ &221.22$\pm$0.44 & 11.23 &   638 & 221.21$\pm$0.50 & 10.65 &   455 & 221.44$\pm$0.94 & 12.65 &   181\\
\enddata
\end{deluxetable*}